\documentclass[sigconf]{acmart}

\AtBeginDocument{%
  }

\copyrightyear{2026}
\acmYear{2026}
\setcopyright{cc}
\setcctype{by}
\acmConference[CHI '26]{Proceedings of the 2026 CHI Conference on Human Factors in Computing Systems}{April 13--17, 2026}{Barcelona, Spain}
\acmBooktitle{Proceedings of the 2026 CHI Conference on Human Factors in Computing Systems (CHI '26), April 13--17, 2026, Barcelona, Spain}
\acmPrice{}
\acmDOI{10.1145/3772318.3790821}
\acmISBN{979-8-4007-2278-3/2026/04}

\usepackage{graphicx}
\newcommand{\Tau}{\scalebox{1.44}{$\tau$}}

\usepackage{fancyvrb}
\usepackage{fvextra}

\usepackage{framed}
\usepackage{xcolor}
\definecolor{shadecolor}{gray}{0.95}

\usepackage{diagbox}
\usepackage{makecell}
\usepackage{booktabs} 
\usepackage{tabularx}
\usepackage{caption}
\usepackage{subcaption}
\usepackage{xurl}

\usepackage{multirow}
\usepackage{multicol}

\usepackage{stfloats}

\usepackage{pifont}%
\newcommand{\cmark}{\textcolor{red}{\ding{51}}}
\newcommand{\xmark}{\textcolor{black}{\ding{55}}}

\begin{document}

\title{Better Assumptions, Stronger Conclusions: The Case for Ordinal Regression in HCI}

\author{Brandon Victor Syiem}
\affiliation{
    \institution{School of Computer Science \\ The University of Sydney}
    \city{Sydney}
    \state{NSW}
    \country{Australia}
}
\email{brandon.syiem@sydney.edu.au}

\author{Eduardo Velloso}
\affiliation{
    \institution{School of Computer Science \\ The University of Sydney}
    \city{Sydney}
    \state{NSW}
    \country{Australia}
}
\email{eduardo.velloso@sydney.edu.au}

\renewcommand{\shortauthors}{Syiem et al.}

\begin{abstract}

Despite the widespread use of ordinal measures in HCI, such as Likert-items, there is little consensus among HCI researchers on the statistical methods used for analysing such data. Both parametric and non-parametric methods have been extensively used within the discipline, with limited reflection on their assumptions and appropriateness for such analyses. In this paper, we examine recent HCI works that report statistical analyses of ordinal measures. We highlight prevalent methods used, discuss their limitations and spotlight key assumptions and oversights that diminish the insights drawn from these methods. Finally, we champion and detail the use of cumulative link (mixed) models (CLM/CLMM) for analysing ordinal data. Further, we provide practical worked examples of applying CLM/CLMMs using R to published open-sourced datasets. This work contributes towards a better understanding of the statistical methods used to analyse ordinal data in HCI and helps to consolidate practices for future work. 

\end{abstract}




\maketitle

\section{Introduction}

Ordinal measures, such as Likert-item surveys, score-based questionnaires, and user preference rankings, are extensively used in HCI research. These measures enable researchers to gather quantitative estimates of users' subjective perceptions, psychological states, attitudes, preferences, judgements, and traits, allowing direct comparative analysis between experimental conditions, otherwise impossible through qualitative data alone~\cite{south2022effective}. Despite the ubiquity and pertinence of such measures in HCI, there is a lack of consensus on the current statistical methods used to analyse such data. This has potential long-term consequences that can hinder HCI researchers from aggregating, contrasting, and replicating findings derived from ordinal data.   

Both parametric and non-parametric statistical methods are frequently encountered for analysing ordinal data in the HCI literature. Parametric methods, such as the one-way repeated ANOVA test, offer more statistical power when compared to their commonly used non-parametric counterparts, such as the Friedman test. However, this increase in statistical power of parametric methods comes at the cost of stronger assumptions, typically on the distribution, ordering, and interval spacing of the data. Previous work has raised significant concerns in analysing ordinal data with parametric methods~\cite{kaptein2010powerful, calver2020anova}. However, these concerns largely stem from the metric (ordered and evenly spaced intervals) handling of ordinal data in widely used methods, as opposed to the method's parametric nature. Analysing ordinal data with metric assumptions has been shown to increase Type I (false positives) and Type II (false negatives) errors, and even inverse the estimated means between groups~\cite{liddell2018analyzing}.

Comparatively, non-parametric methods impose fewer assumptions on data distribution and interval spacing. Importantly, \textit{most} non-parametric methods do not treat ordinal data as metric, and are therefore seen as appropriate for analysing such data~\cite{kaptein2010powerful, jamieson2004likert, schrum2023concerning}. However, non-parametric methods are less sensitive, and could prevent researchers from detecting existing effects~\cite{siegel1957nonparametric}. Moreover, ongoing debate regarding the metric/non-metric nature of data \textit{derived} from ordinal measures, such as aggregated Likert scales~\cite{carifio2008resolving,norman2010likert}, further complicate the selection of appropriate methods. These challenges are particularly salient in HCI, where limited discourse~\cite{kaptein2010powerful, south2022effective} and the conflation of concepts such as `metric' and `parametric' have produced inconsistencies in the statistical analysis of ordinal data.

Such methodological disagreement in HCI research can lead to the propagation of erroneous insights, prevent methodologically distinct but related studies from being cross-examined, and hinder the process of deriving theory from empirical results. To address these issues, we survey current HCI literature to better understand the different statistical methods used for analysing ordinal data. We discuss the underlying mechanisms and assumptions used by frequently used methods in HCI, and deliberate on their appropriateness for handling ordinal data. We then spotlight and promote the use of cumulative link (mixed) models, or CL(M)Ms, that appropriately treat ordinal data as categorical while leveraging the inherent ordering between the categories. We explain the intuition and theory behind CL(M)Ms and detail its use and interpretation through a practical demonstration on open-sourced HCI datasets using R.

We show that despite past criticisms of analysing ordinal data with metric assumptions, they continue to be frequently used in HCI research. This includes frequently used parametric tests with known metric assumptions, as well as commonly used non-parametric methods often regarded as appropriate for ordinal data. Our findings and discussion serve to better inform HCI researchers of the assumptions and pitfalls in current practices employed for statistically analysing ordinal data. Finally, our description of CL(M)Ms and demonstration of its application using R on past open-sourced HCI data provide researchers with the information needed to apply and interpret such methods --- previously considered overly complex and inaccessible to HCI researchers~\cite{wobbrock2011the}.

\section{Background}
\label{sec:related_work}

\subsection{Ordinal Measures, Likert-scales, and HCI} %

Ordinal measures are \textbf{categorical data} that \textbf{have inherent ordering}. Take, for example, a survey item that asks participants to rate how satisfied they were with using a new image editing software on a scale of 1 to 5, with 1 being `Very dissatisfied' and 5 being `Very satisfied'. The participant's response --- a whole number score between 1 and 5 --- is an ordinal measure that clearly indicates better user satisfaction as the rating goes up. However, ordinal data \textbf{carry no metric information}. This means that the distance between ordinal levels may not be equal, and the labels are \textit{purely categorical with known ordering}. As such, the levels 1 to 5 in our example can be replaced by any other set of ordered categories, such as A to E or non-contiguous numerical labels (e.g., 1,2,7,16, and 41), without any loss of information. However, this also suggests that typical statistical methods that involve arithmetic operations to analyse or describe data, such as the data average, cannot be meaningfully interpreted when applied to ordinal data~\cite{marcus1987meaningless}.

Despite such inhospitable properties to arithmetic operations, contiguous numerical labels continue to be the most frequently used labels for representing different levels of ordinal data. This numerical representation often leads ordinal data to be interpreted as an interval or ratio scale. However, the appropriateness of treating ordinal data, and data derived from ordinal measures, as metric has stirred considerable debate across multiple scientific disciplines~\cite{kaptein2010powerful, liddell2018analyzing, marcus1987meaningless, south2022effective, carifio2008resolving} that has spanned nearly a century~\cite{silan2025can}. %

This has not deterred the use of experimental methods that generate ordinal data in scientific disciplines. HCI research, for example, widely makes use of ordinal measures in Likert-scales, surveys and questionnaires, such as the NASA-TLX~\cite{hart1988development} or the system usability scale (SUS)~\cite{brooke1996sus}. Likert scales for instance, consist of multiple related survey items (or Likert items), each with an ordinal response. Their widespread use in multiple disciplines, including HCI, psychology, economics, and medicine, among others~\cite{liddell2018analyzing}, places them at the heart of the dispute surrounding ordinal data analysis. Despite these challenges, Likert scales are invaluable for collecting quantitative estimates of subjective phenomena otherwise unobtainable through qualitative data alone~\cite{south2022effective}. %

Likert scales present an additional dilemma to researchers, i.e., to analyse the scale as a single entity or to analyse individual Likert items. The prescribed method for analysing responses to Likert scales, as per Likert himself~\cite{likert1932technique}, was to aggregate (sum or average) responses to individual Likert items prior to analysis. This treats the data as metric, imposing assumptions so as to meaningfully apply arithmetic operations for analysis. Numerous prior works, in various disciplines, have criticised this approach~\cite{jamieson2004likert, liddell2018analyzing, marcus1987meaningless}, stating that averaging ordinal data cannot produce interval scaled values. Others, however, have argued that aggregated ordinal data is perfectly suitable for metric interpretations, and can be analysed using metric models given their robustness~\cite{carifio2008resolving, carifio2007ten, schrum2023concerning}. 

In contrast, researchers largely agree that individual Likert items, and Likert-type items~\cite{kaptein2010powerful} (single survey items with ordinal responses that are independent of Likert scales), should not be statistically analysed using metric models~\cite{carifio2008resolving, kaptein2010powerful, carifio2007ten}. ~\citet{liddell2018analyzing} present strong arguments against treating averaged Likert scales and Likert-type items as metric, and demonstrate that the intuition behind the interval interpretation of averaged ordinal data is wrong. These disputes are ongoing, and we refrain from prescribing any single interpretation. Instead, we highlight these disputes to demonstrate the extent and persistence of the disagreement surrounding ordinal data analysis, which continues to hinder analytical consistency and reproducibility within HCI.

Additionally, HCI research also frequently employs methods other than Likert scales and Likert-type items that generate ordinal data~\cite{dragicevic2019increasing}. A less obvious example of data interpretable as ordinal in HCI is that of a pre-/post-experiment test on topic understanding that results in a numerical score. While such a test score within a bounded numerical range (say range represented by $R$) intuitively seems to be metric, there is no guarantee that a score increase from $x$ to $x+1$ represents the same level of increase in understanding as a score increase from $y$ to $y+1$ (where $\{x,y\}\  \epsilon \ R$), as this would imply that all questions are equally difficult for all students. The extensive use of ordinal measures in HCI, combined with the inconclusive disputes surrounding its treatment, warrants a deeper reflection on the recent methods used to analyse ordinal data in HCI research. While past work has provided insights into specific HCI subtopics~\cite {south2022effective} and suggestions for analysing ordinal data in HCI~\cite {kaptein2010powerful,wobbrock2011the}, there has been little reflection on the frequency of use of ordinal measures, the methods used to analyse them, and methodological consensus within the field.

\subsection{The Parametric versus Non-parametric Debate}

While concerns about ordinal data analysis primarily stem from metric assumptions, the widespread use of parametric methods that impose these assumptions has led prior work to discuss metric and parametric concepts interchangeably. Consequently, much of the debate surrounding ordinal data has regrettably shifted towards the appropriateness of parametric and non-parametric methods for analysis. For instance, commonly used parametric models in HCI, such as ANOVA and Student's t-test, place assumptions on the distribution of the data to enable meaningful statements to be made around arithmetic estimates, such as the mean and standard deviation. This makes such models appropriate when data can be interpreted as interval or ratio scale. In addition, these methods are not invariant to monotone transformations, i.e., transformations that change the absolute magnitude of the data while preserving relative order. This makes them unsuitable for analysing ordinal data, as ordinal data contain no metric information, and any changes to the magnitude of scale labels should not affect analysis results.

In contrast, most non-parametric models do not impose similar assumptions on data distribution, and are invariant to monotonic transformations. Instead, non-parametric models frequently encountered in HCI, such as the Friedman test, transform the data into relative ordered ranks to further process for insights. For example, the ordered data sets of \{5, 2, 3\} (ordinal levels: 1 - 7) and \{D, B, C\} (ordered levels: A - G), will be ranked as \{3, 1, 2\}. This makes non-parametric models agnostic to changes in magnitude as well; distinct data groups containing \{7, 1, 2\} and \{5, 1, 2\} will similarly be ranked as \{3, 1, 2\} regardless of how much greater 7 is to 5 in the original data. Consequently, these methods are less sensitive to detecting effects when compared to their parametric counterparts, as the former ignores differences in magnitude entirely. However, they are considered appropriate for ordinal data analysis as they impose less stringent assumptions~\cite{south2022effective}.

In addition to sensitivity concerns, there is also a lack of accessible non-parametric methods for analysing multi-factorial experimental designs for HCI researchers. Prior work has addressed this challenge by highlighting existing methods and providing access to tools that enable non-parametric data analysis for multi-factorial designs~\cite{wobbrock2011the, kaptein2010powerful}. For instance,~\citet{kaptein2010powerful} highlights existing non-parametric methods used in the medical field for analysing Likert-type item data. Similarly,~\citet{wobbrock2011the} introduces the aligned ranked transform (ART) method that enables practitioners to apply familiar and powerful multi-factorial ANOVA tests on transformed non-parametric data. Despite these efforts, there is still little consensus within the HCI discipline on how to best analyse ordinal data~\cite{dragicevic2019increasing}.

A largely neglected concern with using non-parametric methods for ordinal data analysis within HCI relates to test-specific assumptions. Specifically, while non-parametric methods can be used to analyse ordinal data without \textit{imposing the normality assumption} (i.e., assuming that the residuals are normally distributed), they may \textit{still assume metric data}. Non-parametric methods, such as the Wilcoxon Rank Sum Test~\footnote{ranks data points across both samples to be compared} or the Friedman Test~\footnote{ranks data points within each repeating block across all levels of a factor}, are examples of tests that do not assume normality of residuals or metric data. Therefore, these methods do not violate any of the properties of ordinal data, but do ignore the difference in distances between ordinal levels, which may reduce statistical power. In contrast, methods such as the Quade Test or the Wilcoxon Signed Rank Test for paired samples first calculate the \textit{paired differences prior to assigning ranks}. This makes these tests unsuitable for strictly ordinal data as they perform arithmetic operations on data with no metric information~\cite{kornbrot1990rank}. Similarly, while it is suggested that the ART method can be used for ordinal data analysis~\cite{wobbrock2011the}, the process of `aligning' or transforming the data in ART requires the calculation of averages and residuals from the unaltered ordinal data, and, as such, imposes metric assumptions on the data that may not be suitable for ordinal data.

Imposing such assumptions is unlikely~\cite{silan2025can} to be problematic by themselves, provided the assumptions and possible drawbacks are made explicit to readers. However, more recent approaches have been developed that are specifically designed to leverage the information held by ordinal data, without imposing incompatible assumptions on the data, such as with the use of cumulative link (mixed) models, or CL(M)Ms. The availability of these methods, along with software and programming tools that make these methods more accessible and practical to use, presents an opportunity to reconcile ordinal data analysis practices in HCI and move closer to more robust, theoretically informed, and reproducible findings.

\section{Review of Ordinal Analysis Practices in HCI}
\label{sec:systematic_review}
To understand current methodological practices for analysing ordinal data in HCI, we conducted a review of recent research articles published in the proceedings of the ACM CHI Conference on Human Factors in Computing Systems (CHI). We chose CHI based on its reputation as the flagship conference for HCI research, and its broader representation of HCI topics (unlike more focused venues, such as ISMAR or Ubicomp).

\subsection{Sampling}

As our aim was to understand \textit{recent} statistical approaches for ordinal data analysis in HCI, we restricted our search to the year 2024 and used a broad search string: \textit{[All Field: (``questionnaire'' OR ``likert'')]}. We also limited our search to full papers. We included papers that conducted user studies and reported statistical analyses of ordinal data. We excluded papers that (1) did not include ordinal data, (2) focused on scale development and not on generating insights from analysing ordinal data, and (3) did not report analyses of included ordinal data beyond summary statistics.

\begin{figure}[tbp]
    \centering
    \includegraphics[width = 1.0\columnwidth]{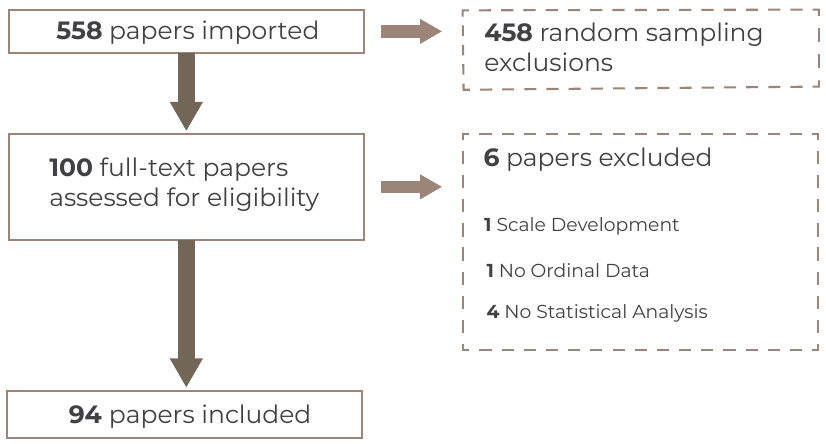}
    \caption{Overview of our review process detailing each step reported as per the PRISMA guidelines~\cite{moher2009preferred}. We randomly sampled 100 papers from our initial search result of 558 papers. The full text of all 100 papers was screened for eligibility. Six (6) papers were excluded, and the remaining 94 papers were included in our final sample.}
    \Description{PRISMA diagram showing the flow of papers through the review process: 558 papers identified, 100 sampled, 6 excluded, 94 included.}
    \label{fig:prisma-diagram}
\end{figure}

Our search resulted in 558 papers. We randomly sampled 100 papers from the search results for full-text analysis. From the 100 sampled papers, we excluded 6 papers, resulting in a final sample of 94 papers (details of the review steps are presented in Figure~\ref{fig:prisma-diagram}). For each paper, we recorded all ordinal measures collected and the associated construct. We then recorded the statistical method used to analyse ordinal measures related to each construct. Note that a single paper could report multiple studies, and each study may collect multiple ordinal measures that may be analysed using multiple statistical methods. All extracted data can be found in our supplemental material.   

\subsection{Findings}

We separate our findings into three categories of statistical analysis methods; (1) statistical methods used to \textit{mathematically represent the data, and can be used for parameter estimation and outcome predictions in addition to hypothesis testing} --- \textbf{\textit{predictive modelling}}, (2) statistical methods used \textit{primarily for hypothesis testing involving multiple factors and/or more than two groups} --- \textbf{\textit{omnibus tests}}, and (3) statistical methods used \textit{primarily for hypothesis tests between two groups, or for performing multiple two group comparisons with corrections}~\footnote{Note that we do not extract data related to the specific corrections used during multiple pairwise comparisons} (such as the Bonferroni correction) --- \textbf{\textit{pairwise comparisons/multiple pairwise comparisons}}.

\begin{figure}[tbp]
    \centering
    \includegraphics[width = 1.0\columnwidth]{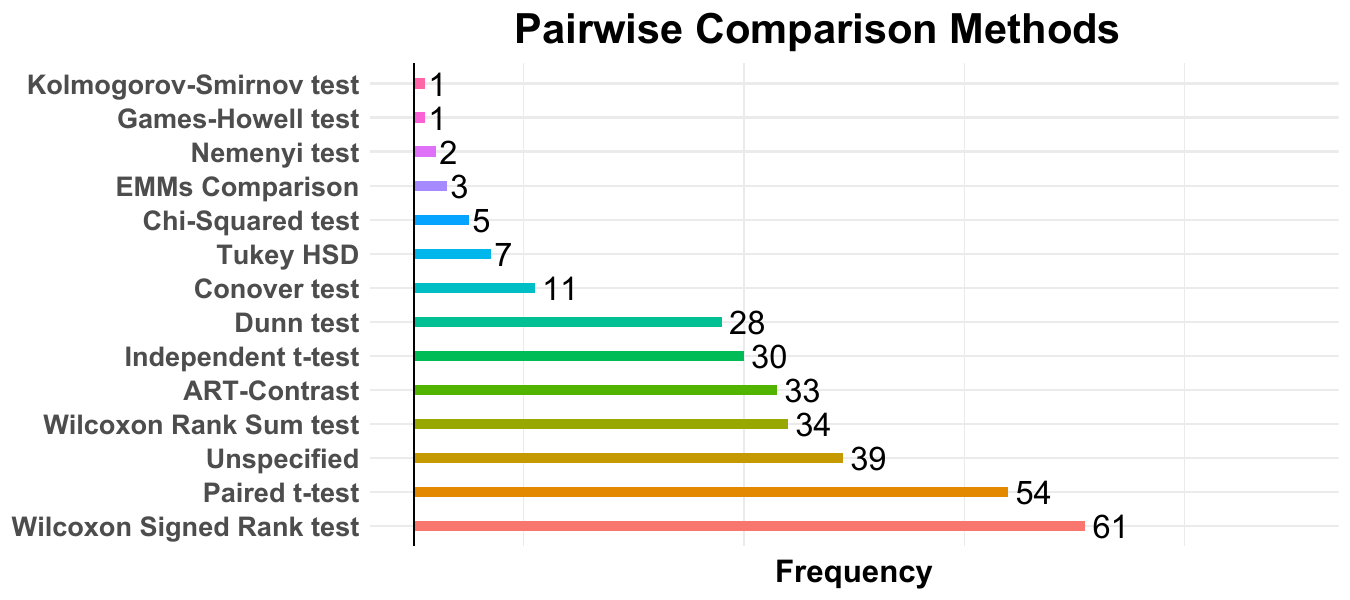}
    \caption{Frequency of different pairwise comparisons reportedly used for ordinal data analysis in our sample.}
    \Description{Histogram showing the frequency of statistical tests used for pairwise comparison in our sample. Eighty-four instances of t-tests (30 independent-sample and 54 paired-sample t-tests), 7 Tukey HSD tests, 3 EMMs Comparisons, 61 Wilcoxon Signed Rank tests, 34 Wilcoxon Rank Sum test, 5 Chi-squared tests, 28 Dunn tests, 11 Conover tests, 1 Games-Howell test, 1 Kolmogorov-Smirnov test, 2 Nemenyi tests, 33 ART-Contrast tests, and 39 tests that were not specified (only referred to as post-hoc tests).}
    \label{fig:pairwise}
\end{figure}

\paragraph{\textbf{Predictive modelling.}} Only 28 instances of predictive modelling were reported for ordinal data analysis in our sample. Figure~\ref{fig:pairwise} shows the frequency of the different reported models used in our sample. Eight analyses used the Cumulative Link Mixed Model (CLMM), 2 used the Cumulative Link Model (CLM), 3 uses of the Generalised Estimation Equation, 1 use of Generalised Linear Mixed Model (with a logit link function --- not cumulative logit which would class it as a CLMM), 9 instances of Linear Mixed-Effects Models, 4 Linear Regression Models, and 1 analysis using the PROCESS model 7~\cite{hayes2012process}.

\begin{figure}[tbp]
    \centering
    \includegraphics[width = 1.0\columnwidth]{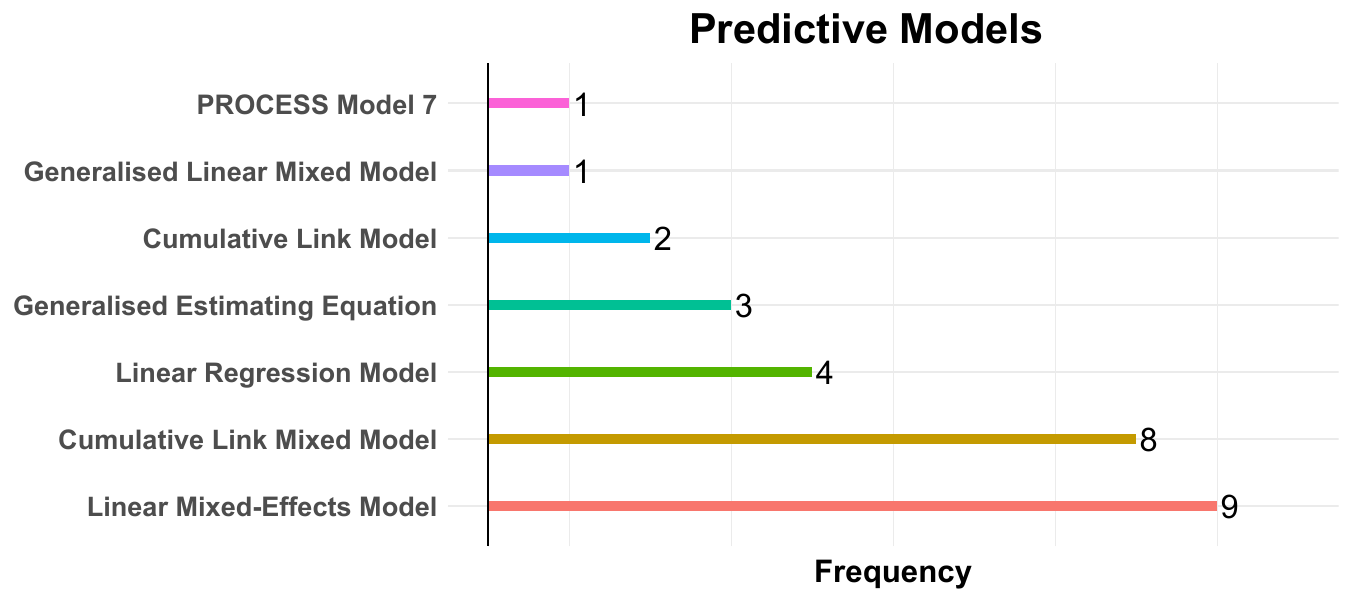}
    \caption{Frequency of different predictive models reportedly used for ordinal data analysis in our sample.}
    \Description{Histogram showing the frequency of predictive models used for ordinal data analysis in our sample. 8 Cumulative Link Mixed Models, 2 Cumulative Link Models, 3 models using Generalised Estimation Equations, 1 Generalised Linear Mixed Model, 9 Linear Mixed-Effects Model, 4 Linear Regression Models, and 1 Hayes PROCESS model 7.}
    \label{fig:predictive_models}
\end{figure}

\begin{figure}[tbp]
    \centering
    \includegraphics[width = 1.0\columnwidth]{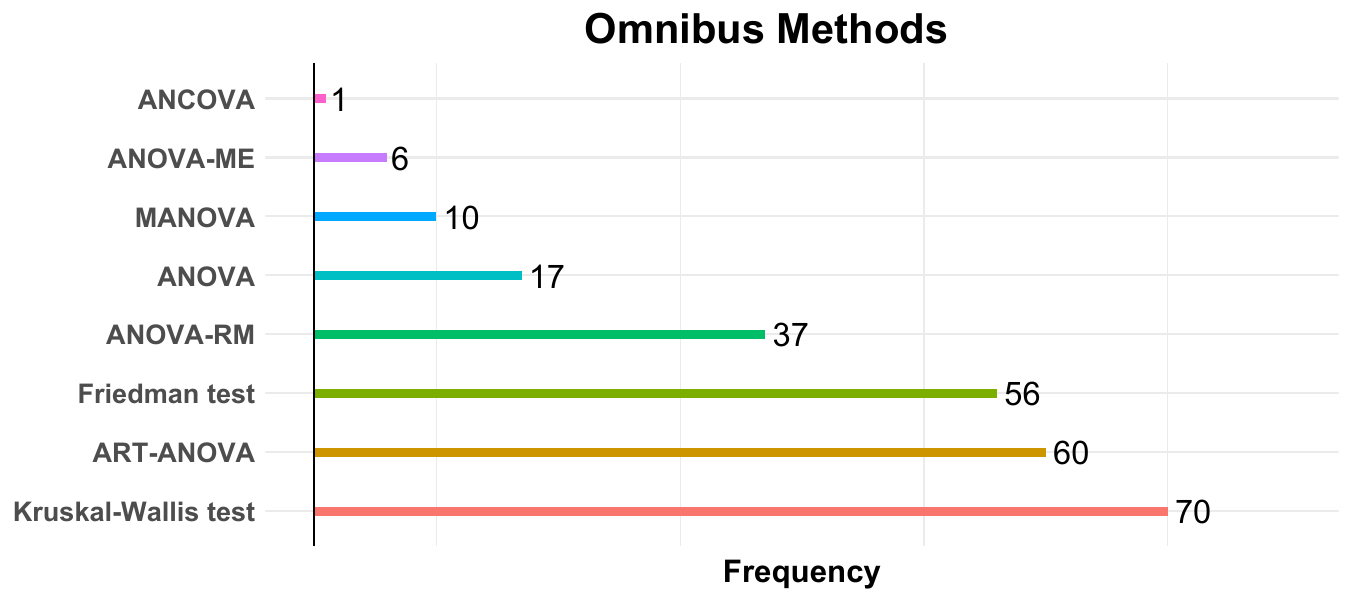}
    \caption{Frequency of different omnibus tests used for ordinal data analysis in our sample.}
    \Description{Histogram showing the frequency of omnibus tests used to analyse ordinal data found in our sample. Seventy-one instances of ANOVA variants (1 ANCOVA, 6 Mixed-effects ANOVA, 10 MANOVA, 17 ANOVA, 37 Repeated Measure ANOVA), sixty instances of ART-ANOVA, fifty-six instances of Friedman tests, and seventy instances of Kruskal-Wallis tests.}
    \label{fig:omnibus}
\end{figure}

\paragraph{\textbf{Omnibus tests.}} A total of 257 omnibus tests were reported to be used for ordinal data analysis in our sample. Figure~\ref{fig:omnibus} shows the frequency of the different reported tests. Parametric methods comprised 71 reported uses of various Analysis of Variance (ANOVA) tests, including 17 independent-sample ANOVA,  37 repeated-measure ANOVA (ANOVA-RM), 6 mixed-effects ANOVA (ANOVA-ME), 1 Analysis of Covariance (ANCOVA), and 10 multivariate ANOVA (MANOVA). Reported non-parametric methods include 60 reported cases of the ART method to transform data prior to ANOVA analysis~\cite{wobbrock2011the} (ART-ANOVA), 70 analyses using the Kruskal-Wallis test, and 56 analyses using the Friedman test.  

\paragraph{\textbf{Pairwise comparisons.}} 309 pairwise comparisons were reportedly conducted for ordinal data analysis in our sample. Figure~\ref{fig:pairwise} shows the frequency of the different reported tests. Parametric tests were dominated by paired (54) and independent-sample t-tests (30), with a minority employing the Tukey Honest Significant Difference (Tukey HSD) test (7). Non-parametric tests primarily consisted of the Wilcoxon Signed Rank test (61) for paired samples, and the Wilcoxon Rank Sum test (34) for independent samples. Other non-parametric tests observed include the Chi-squared test (5), the Dunn test (28), the Conover test (11), Games-Howell test (1), the Kolmogorov-Smirnov test (1), the Nemenyi test (2), and the ART-Contrast tests (33) following the ART-ANOVA tests. Additionally, we found three (3) instances of pairwise comparisons based on estimated marginal means (EMMs Comparison). Note that `EMMs Comparison' is not a single statistical test, but refers to post-hoc pairwise comparisons of estimated marginal means derived from a fitted model~\cite{lenth2025emmeans}---observed following 1 GLMM and 2 CLM model fits in our sample~\footnote{We classify `EMMs Comparison' as parametric because estimated marginal means depend on model parameters. However, it can also be applied in non-parametric contexts, such as post-hoc comparisons following an ART-ANOVA model fit}. Finally, we found multiple post-hoc tests in our sample that did not specify the exact test used and simply referred to pairwise comparisons as `post-hoc analysis/test'. This accounted for 39 instances in our sample, labelled as `unspecified'.

\begin{figure*}[tbp]
    \centering
    \includegraphics[width = 1.0\textwidth]{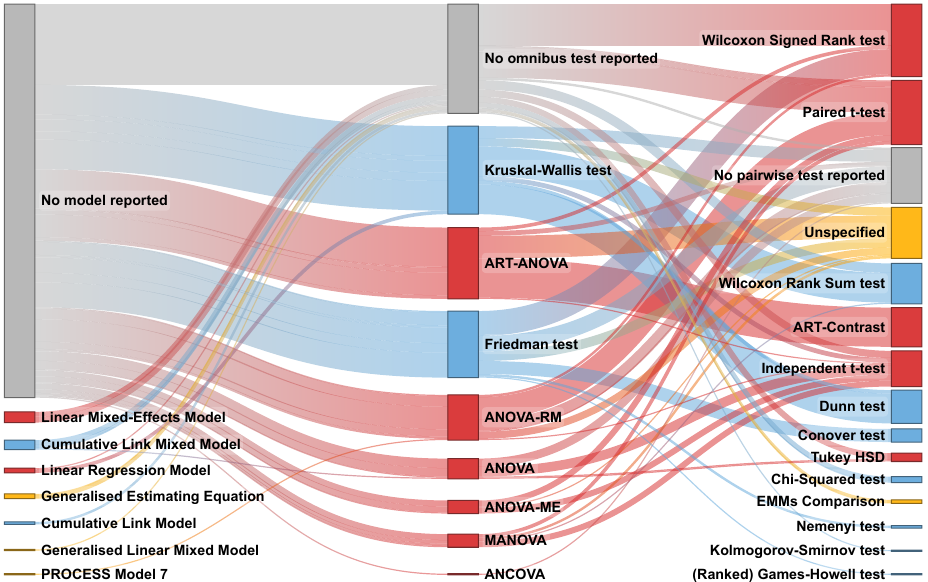}
    \caption{Sankey diagram showing the progression of statistical procedures applied to constructs measured with ordinal data. Methods are displayed in sequence from predictive models (left), to omnibus tests (centre), to pairwise tests (right). Bars labelled ‘No’ indicate constructs for which no method was reported in that category. Red nodes indicate statistical methods that impose incompatible assumptions with ordinal data, blue nodes represent ordinal data compatible methods, yellow nodes represent methods whose appropriateness for analysing ordinal data is dependent on specific factors (such as the link' function used), and grey nodes represent absent methods in that specific category. As seen in the Figure, our sample reported limited use of predictive models (No model reported), and most models and omnibus approaches were not frequently used together (most models have links to `No omnibus test reported'). The Figure also clearly highlights a lack of consensus within the field; with no clear direction of which method, or sequence of methods, should be used to analyse ordinal data.}
    \Description{Sankey Diagram depicting the sequence of statistical approaches used to analyse ordinal data in our sample; starting from predictive models, to omnibus tests and finally to pairwise comparisons. Red nodes indicate statistical methods that impose incompatible assumptions with ordinal data, blue nodes represent ordinal data compatible methods, yellow nodes represent methods whose appropriateness for analysing ordinal data is dependent on specific factors (such as the link' function used), and grey nodes represent absent methods in that specific category. The Figure depicts a lack of use of predictive models to analyse ordinal data in our sample. Additionally, the high number of branching links between the categories (model/omnibus/pairwise) suggests a lack of consensus in the field.}
    \label{fig:sankey_statistical_procedures}
\end{figure*}

\subsection{Discussion}
\label{sec:systematic_review_discussion}

\paragraph{\textbf{An undecided discipline:}} Our review demonstrates the use of numerous statistical approaches for analysing ordinal data in HCI research. Both parametric and non-parametric methods continue to be frequently used, with non-parametric approaches outnumbering parametric approaches (362~\footnote{60 ART-ANOVA + 70 Kruskal-Wallis tests + 56 Friedman tests + 33 ART-Contrast + 5 Chi-squared test + 28 Dunn test + 11 Conover test + 1 Games-Howell test + 1 Kolmogorov-Smirnov test + 2 Nemenyi test + 34 Wilcoxon Rank Sum test + 61 Wilcoxon Signed Rank test} against 165\footnote{17 ANOVA + 37 ANOVA.RM + 6 ANOVA.ME + 1 ANCOVA + 10 MANOVA + 3 EMMEANS + 30 independent t-test + 54 paired t-test + 7 Tukey HSD} respectively, excluding unspecified pairwise comparisons and predictive models; as models are dependent on the link function used). The increased use of non-parametric methods suggests a trend towards adopting approaches that impose fewer assumptions on ordinal data for analysis. However, we observed 154 cases of the use of the non-parametric approaches that assume data to be metric. These include 60 cases of the ART-ANOVA test, 33 cases of the post-hoc ART-Contrasts test, and 61 cases of the Wilcoxon signed rank test. This suggests that a total of 319 out of 527 tests in our sample (excluding models and unspecified pairwise comparisons) impose metric assumptions on ordinal data. It is unclear if this is the result of widespread misinterpretation of the term `non-parametric' to also \textit{exclude metric assumptions on the data} or an informed \textit{decision to treat ordinal data as metric}. In either case, this presents a challenge for future scholars in consistently interpreting, contrasting and building upon findings related to ordinal data that rely on different data assumptions. To better inform practitioners, we present the relevant assumptions for ordinal data analysis of the different hypothesis tests (omnibus and pairwise comparisons) found in our sample in Table~\ref{tab:nonparametric_test_assumptions}. The table excludes predictive models as their assumptions are dependent on additional factors, such as the `link' function used. %

\begin{table*}[tbph]
    \resizebox{\textwidth}{!}{
    \centering
    \begin{tabular}{lccccc}
        \toprule
         \multirow{2}{*}{\textbf{Omnibus or Pairwise Method}} & \multirow{2}{*}{\textbf{Category}} & \multirow{2}{*}{\textbf{Experimental Design}} & \multicolumn{3}{c}{\textbf{Assumption}} \\ \cline{4 - 6}
        &  & &\textbf{Metric Data} & \textbf{Normality of Residuals} & \textbf{Equal Variance} \\
         \midrule
         Wilcoxon Signed Rank test~\cite{wilcoxon1945individual, kornbrot1990rank} & Pairwise & Within-Subject Design &\cmark & \xmark & \xmark \\
         Wilcoxon Rank Sum test~\cite{wilcoxon1945individual} & Pairwise & Between-Subject Design &\xmark & \xmark & \xmark \\
         ART-Contrasts~\cite{elkin2021aligned} & Pairwise & Mixed Design & \cmark & \xmark & \xmark \\
         Dunn test~\cite{dunn1964multiple,dinno2015nonparametric} & Pairwise & Between-Subject Design & \xmark & \xmark & \xmark \\
         Conover test~\cite{conover1979multiple,conover1999practical,pohlert2024the} & Pairwise & Between/Within-Subject Design & \xmark & \xmark & \xmark \\
         Chi-squared test~\cite{pearson1900x} & Pairwise & Between-Subject Design & \xmark & \xmark & \xmark \\
         Nemenyi test~\cite{nemenyi1963distribution} & Pairwise & Within-Subject Design & \xmark & \xmark & \xmark \\
         (Ranked) Games-Howell test~\cite{games1976pairwise, shingala2015comparison} & Pairwise & Between-Subject Design & \xmark & \xmark & \xmark \\
         Kolmogorov-Smirnov test~\cite{berger2014kolmogorov} & Pairwise & Between-Subject Design & \xmark & \xmark & \xmark \\
         Kruskal-Wallis test~\cite{kruskal1952use} & Omnibus & Between-Subject Design & \xmark & \xmark & \xmark \\
         ART-ANOVA~\cite{wobbrock2011the} & Omnibus & Mixed Design & \cmark & \xmark & \xmark \\
         Friedman test~\cite{friedman1937use} & Omnibus & Within-Subject Design & \xmark & \xmark & \xmark \\
         Sampled Parametric Methods & --- & --- & \cmark & \cmark & \cmark \\
         \bottomrule
    \end{tabular}
    }
    \caption{Table detailing common assumptions relevant to ordinal data analysis for omnibus and pairwise methods found in our sample. The table lists the category of the statistical method, the experimental design that generates data suitable for the method to analyse, and the assumptions that the method imposes. The table primarily details sampled non-parametric methods, as all parametric omnibus \& pairwise tests in our sample (not all existing parametric methods) assume metric data, normality of residues, and equal variance --- and are hence collectively represented by the last row. Notably, this table highlights frequently used non-parametric methods in HCI that exhibit incompatible assumptions on ordinal data. Specifically, the Wilcoxon Signed Rank test, the ART-Contrasts and the ART-ANOVA impose metric assumptions on the data (via the use of arithmetic operations on unranked data). Note that this table excludes approaches that we categorize as predictive models (e.g., CLMs or GLMs), as their assumptions depend on additional factors such as the `link’ function.}
    \label{tab:nonparametric_test_assumptions}
\end{table*}

\paragraph{\textbf{Inconsistent successive assumptions:}} Our findings further reveal that current HCI practices for analysing ordinal data may impose inconsistent data assumptions across the sequence of statistical procedures applied to ordinal data within the same study. Take, for example, the pairwise tests following a Friedman test (appropriate for repeated-measure non-parametric data) in our sample as depicted in Figure~\ref{fig:sankey_statistical_procedures}. This includes the Wilcoxon Signed Rank test, the Conover test, the Games-Howell test and the Nemenyi test, among additional unspecified tests. These tests are all appropriate for analysing non-parametric data generated through a within-subject experimental design. However, they differ in their treatment of the data, and may unintentionally impose assumptions incompatible with ordinal data that was initially avoided in the preceding omnibus test. For instance, the most frequently used test following the Friedman omnibus test in our sample was the \textit{Wilcoxon Signed Rank test, which assumes metric data}, unlike its preceding omnibus test (see Table~\ref{tab:nonparametric_test_assumptions}). This inconsistent treatment of ordinal data within individual sequences of analyses further complicates the interpretation of ordinal data analysis, which is already challenged by the parametric and non-parametric debate. Such inconsistencies warrant a change in the current statistical methods used for ordinal data analysis within HCI, prompting the need for methods that address the concerns stemming from over- or under-estimating the information present in ordinal data~\cite{burkner2019ordinal}. 

\paragraph{\textbf{Ordinal first modelling:}} Specifically designed predictive modeling techniques, known as ordinal regression models (e.g., CL(M)Ms), provide greater statistical power than non-parametric approaches while avoiding the incompatible assumptions of commonly used parametric methods in HCI, such as ANOVAs. They achieve this by leveraging the ordinal structure of the data without assuming equally spaced intervals between categories. However, such models see limited application in HCI (see Figure~\ref{fig:predictive_models}). A possible explanation for such limited use, is the focus on the use of ordinal data for hypothesis testing in recent HCI work, as corroborated by the disparity between the frequency of hypothesis tests (omnibus and pairwise tests) and predictive modelling methods in our sample (Figures~\ref{fig:omnibus},~\ref{fig:pairwise}, and~\ref{fig:predictive_models}). This could further be explained by previous challenges in understanding, interpreting and applying predictive modelling methods in HCI research~\cite{wobbrock2011the}. However, we argue that recent developments in statistical analysis software have made these methods increasingly more accessible and easy to apply. To better inform HCI practitioners in understanding and using the methods, the following sections describe the proposed CL(M)M modelling technique, and presents worked examples using CL(M)Ms to analyse ordinal data in published open source HCI data. We additionally provide an interactive tool in our supplementary material to illustrate a simple CLM model.

\section{Cumulative Link (Mixed) Models}
\label{sec:clmm_theory}

Cumulative Link (Mixed) Models, or CL(M)Ms, are a class of Generalised Linear (Mixed) Models that are designed for ordinal regression. Throughout this paper, we will often use the term `CL(M)M' to refer to both cumulative link models (CLMs) and cumulative link mixed models (CLMMs). \textcolor{black}{Specifically, CLMs are appropriate when only fixed effects are of interest and observations can be assumed to be independent. In contrast, CLMMs extend CLMs by incorporating random effects, allowing them to account for grouped or hierarchical data in which the independence assumption is violated (e.g., repeated-measures experimental designs).} These models appropriately treat ordinal data as \textit{categorical}, while \textit{exploiting the ordered nature} of the data~\cite{christensen2018cumulative}. Given such advantages, these methods are increasingly encouraged and used for modelling ordinal data in diverse fields of research, including psychology~\cite{liddell2018analyzing,burkner2019ordinal, gambarota2024ordinal}, econometrics~\cite{katchova2020econometrics}, machine learning~\cite{williams2025cumulative}, and health~\cite{chen2020examining}. Despite the advantages of CL(M)Ms, they have been largely overlooked in HCI, due to challenges in understanding, interpreting and applying these methods~\cite{wobbrock2011the}. As with all statistical methods, CL(M)Ms come with their own set of assumptions that are important to understand and consider prior to their use. With the increasing availability of software packages capable of performing ordinal regression using CL(M)Ms~\cite{christensen2018cumulative, aguena2021clmm} in popular statistical tools, such as Stata, R, and Python, this section provides an intuitive description of the theory and assumptions of CL(M)Ms in the context of HCI research. \textcolor{black}{Readers already familiar with CL(M)M theory and assumptions, or those interested primarily in practical applications of CL(M)Ms for HCI data, may skip to Section~\ref{sec:analysing_using_clmms}.}

\subsection{Intuition}
\label{sec:clmm_intuition}

\begin{figure}[h]
    \centering
    \begin{minipage}[t]{0.49\columnwidth}
        \centering
        \includegraphics[width=0.8\columnwidth]{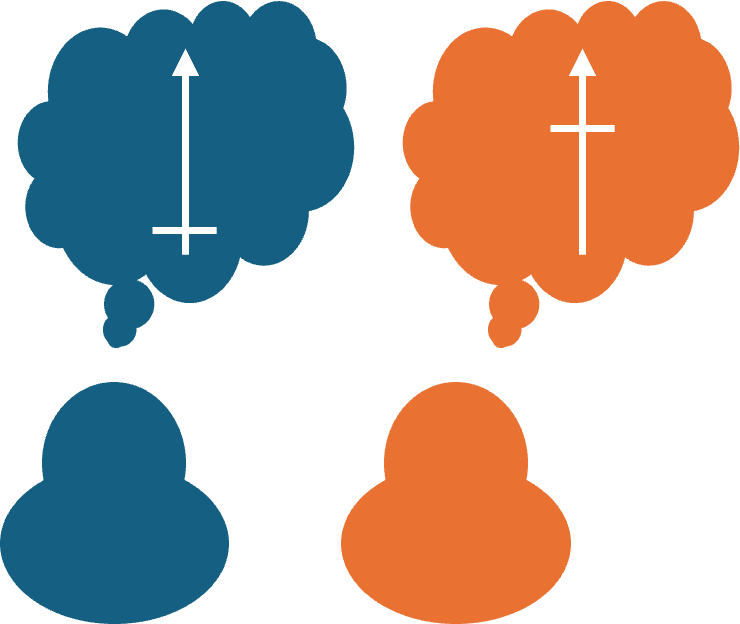}
    \end{minipage}
    \begin{minipage}[t]{0.49\columnwidth}
        \centering
        \includegraphics[width=1\columnwidth]{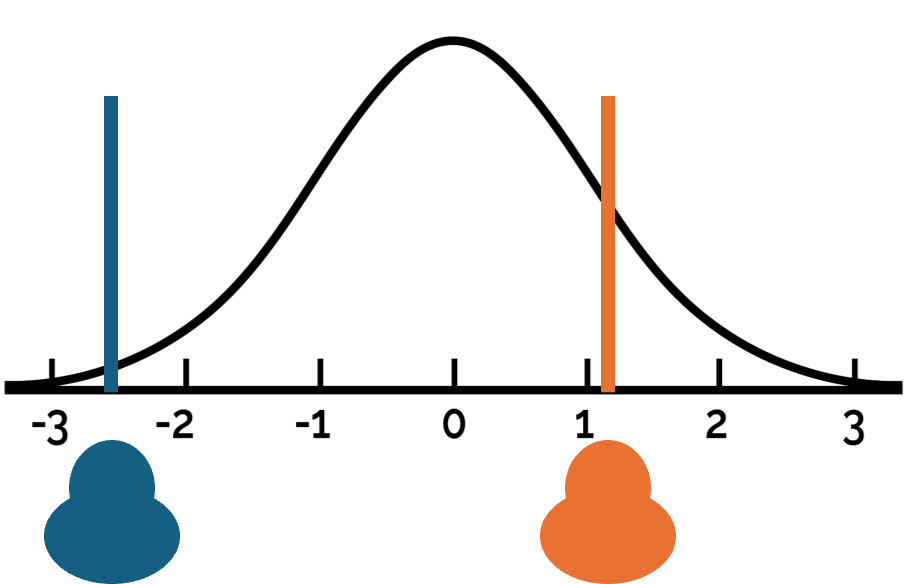}
    \end{minipage}
    \caption{Ideally, we could directly measure a user's sense of spatial presence from their thoughts (LEFT). If the measure is continuous and normally distributed (RIGHT), it would enable us to use familiar metric models of statistical analysis and comparisons.}
    \Description{Illustration of an ideal scenario where users' sense of spatial presence can be directly measured from the thoughts. The figure further highlights how familiar metric models could be applied to such a directly observable measure if it was continuous and normally distributed.}
    \label{fig:ideal}
\end{figure}

Say that we are interested in the effect of adding a new feature to our system (e.g. faster screen frame rate) on an element of user experience (e.g. spatial presence in virtual reality). In an ideal scenario, we would have a tool that could directly measure a user's sense of spatial presence, giving us a continuous and normally distributed measure that we could analyse using a familiar method, such as an ANOVA. Figure~\ref{fig:ideal} illustrates this ideal scenario.

\begin{figure}[h!]
    \centering
    \includegraphics[width=\columnwidth]{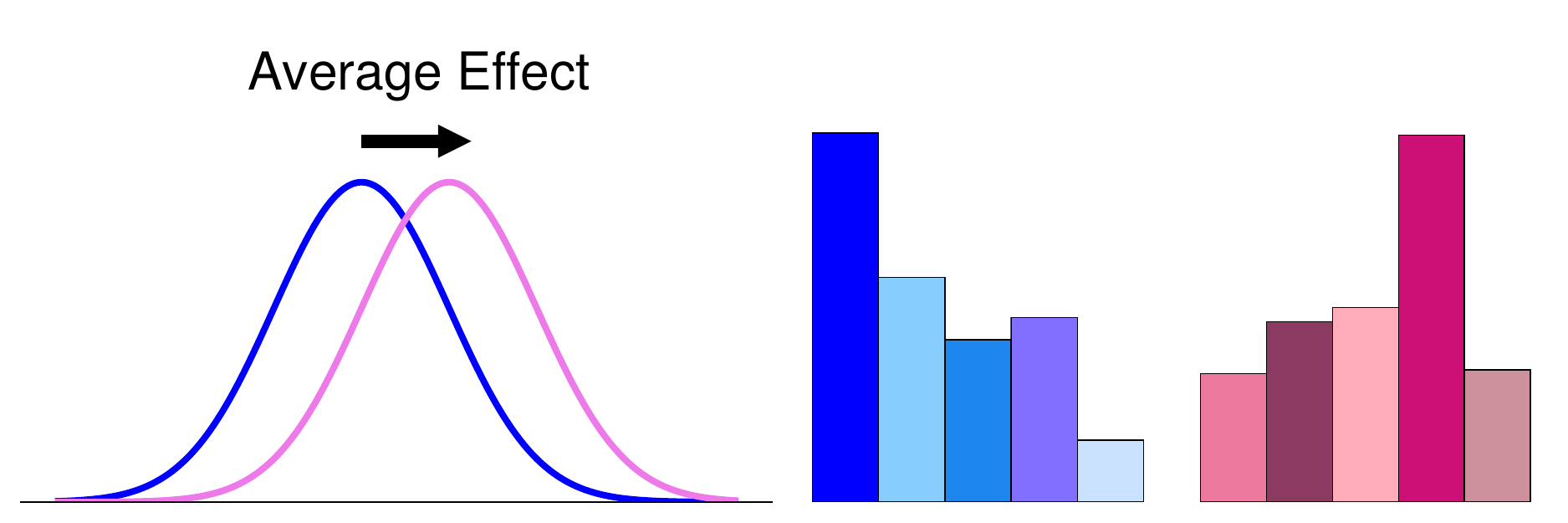}
    \caption{Figure depicting an ideal scenario and our reality when deriving insights from ordinal data related to our system. In an ideal scenario (LEFT), we would not need ordinal measures, and we would have direct and continuous measures of users' spatial presence when using the two versions of our system. This would result in two distinct normal distributions of users' spatial presence when using our system with and without the new feature. In reality (RIGHT), spatial presence cannot be measured directly and instead a discrete score (e.g., between 1 and 5) is provided by the user to indicate their level of spatial presence under the two system variations.}
    \Description{Two images (left and right). Left image depicts an ideal situation where users' sense of spatial presence across two conditions can be measured as a continuous measure, leading to two distinct normal distributions. Right image depicts a real scenario where spatial presence is directly unobservable and can only be measures using ordinal ratings, leading to two ordinal distributions (histograms).}
    \label{fig:step3}
\end{figure}

However, constructs such as spatial presence are not directly observable, and consequently, require us to use proxy measures. Often, Likert-item styled questionnaires are used, where users provide a rating of spatial presence on an ordinal scale. Unlike metric data, ordinal data do not exhibit equal distances between ratings, i.e., the ordinal values of $3 - 2$ and $5 - 4$ may not be equal, even if they seem to arithmetically equate to 1 due to the use of contiguous numerical labels. As such, ordinal values cannot be meaningfully analysed using methods that rely on arithmetic operations. In the context of our example, this means that we cannot estimate how much more spatially present users were (on average) when using our system with, and without, the new feature by subtracting the average ordinal responses for the different system conditions. Figure~\ref{fig:step3} illustrates these differences between the continuous representation in our ideal case (left), and the ordinal measure collected in a real-world context (right). In the ideal scenario, users’ spatial presence scores with and without the added feature form two normal distributions, and the average effect of the new feature is reflected by the shift from the original distribution. In reality, however, we can only collect user spatial presence measures via ordinal ratings on a predefined scale (say from 1 to 5), resulting in distributions of ordinal data (right) that is far from normal, and whose effect is much harder to estimate.

CLMMs attempt to address this challenge by taking a principled approach to analysing ordinal data. These models assume that there exists a continuous \textit{unobservable} measure (or latent variable) in the user's mind that underlies an observable ordinal measure. Since this measure is unobservable, the unit of measurement can be arbitrary. Conventionally, CL(M)Ms assume that this latent variable is distributed with a standard normal (0,1) distribution (though models can be extended into any arbitrary latent distribution). If indeed the observed scores reflect these assumed underlying latent values, then they should be statistically related. However, CL(M)Ms do not assume this relationship to be linear, only monotonic (i.e. as the latent value increases, so does the likelihood of a higher user rating). Imagine a user who initially experiences no spatial presence (i.e. their latent score should be the minimum imaginable). This user would most likely give the lowest possible rating (1 - Very low). As we intervene to increase the latent score from its absolute minimum, we would still see scores of 1 until we reach a certain level, at which point, we would start seeing scores of 2. As we keep increasing the latent score and reaching these thresholds/cutpoints, our ratings should increase accordingly.

\begin{figure}
    \centering
    \includegraphics[width=\columnwidth]{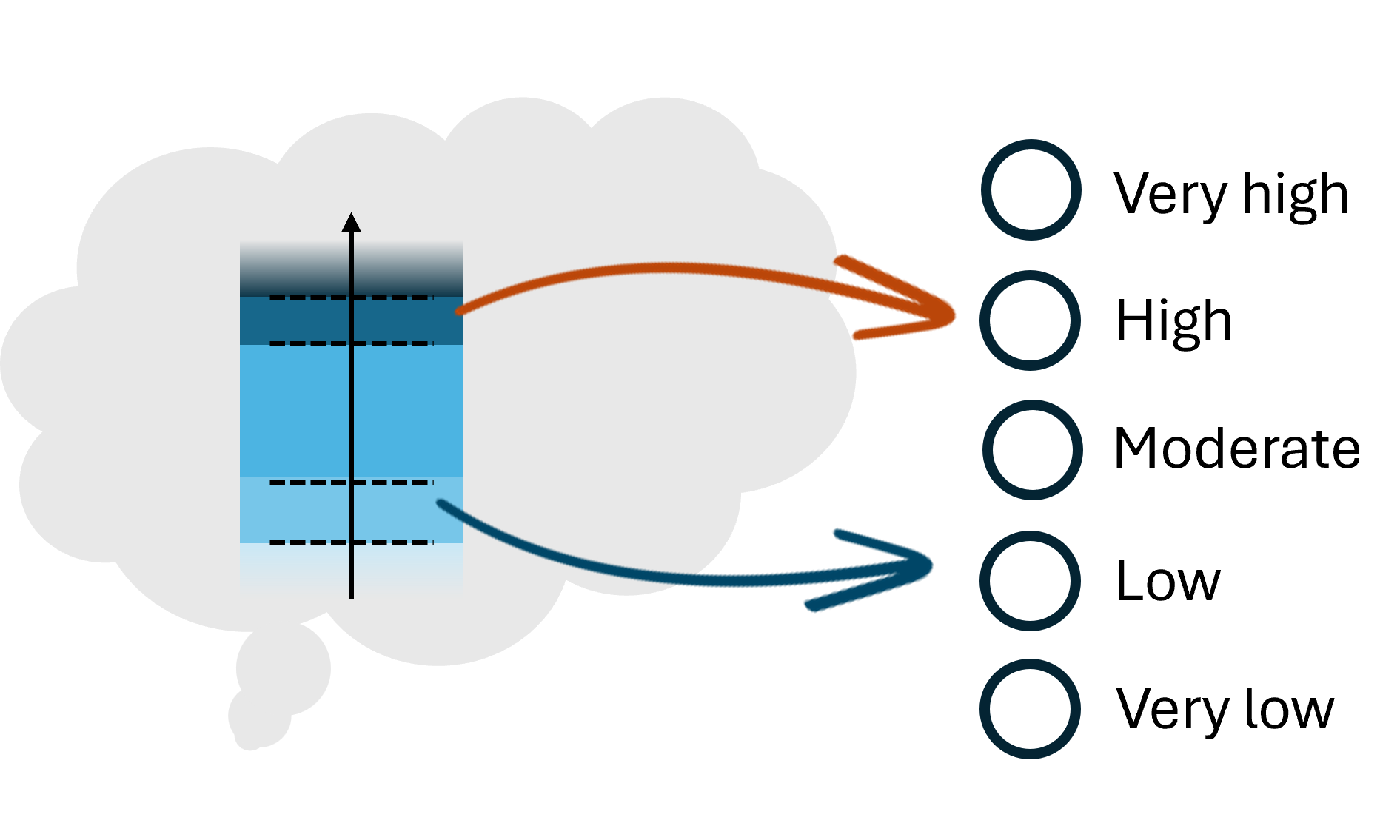}
    \caption{Our ordinal rating increases when the latent value increases beyond a cutpoint. By looking at which cutpoints the latent value falls between, we can predict which score the user would give.}
    \Description{Illustration of how different cutpoints divide the latent variable scale to determine the ordinal response of a user on a Likert item question. As the value of the latent variable increases beyond a cutpoints, the ordinal user response also increases.}
    \label{fig:step4}
\end{figure}

Because the latent normal and the observed scores distributions are connected under these assumptions, we can also assume that the proportion of users with a latent value between any 2 cutpoints should be equal to the proportion of users that gave the corresponding rating. As such, we can compare the cumulative distribution of the scores to the cumulative distribution of the latent normal. In other words, if 10\% of our scores were 1, and we know that in a standard normal distribution 10\% of values are lower than -1.28, then we can say that -1.28 is the cutpoint in the latent scale at which someone's rating would go from 1 to 2. If in the same dataset we found that 15\% of scores were 2, we would have to look for the 15 + 10 = 25\% quantile of the normal distribution to find the next cutpoint---in this case, -0.67. We can proceed the same way for the other cutpoints. Figure~\ref{fig:step4} visualizes the selection of ordinal scores based on the cutpoints that divide the latent measure scale. 

This method gives us a way to retrieve the latent cutpoints from observed scores. However, ultimately, we are interested in the causal effect of our intervention. For this example, we will assume that the intervention only shifts the mean of our latent distribution (not the variance). Under the same cutpoints, shifting the mean of the latent measure causes the area under the normal curve between the cutpoints to change---and these areas are precisely the proportion of each of the possible ratings.

\begin{figure}
    \centering
    \includegraphics[width=1\linewidth]{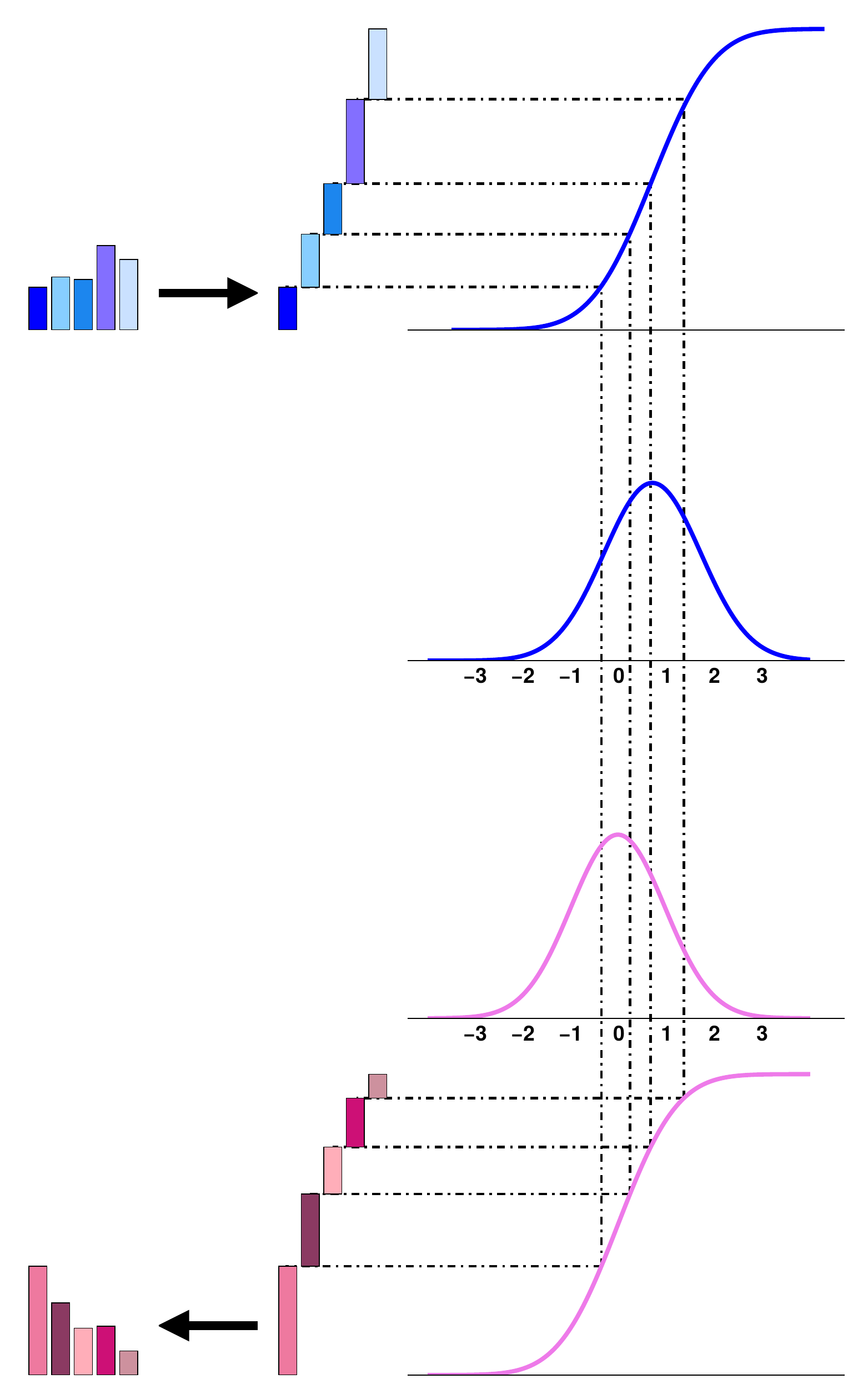}
    \caption{From a given distribution of baseline scores, we can infer the latent cutpoints and predict how shifting the mean of the distribution would affect the distribution of scores accordingly.}
    \Description{Image illustrating the process behind CL(M)Ms. The images shows different distributions related to ordinal ratings for two conditions (baseline and experimental). Each condition has a ordinal distribution (histogram) for the observable user response, and two distributions (normal and cumulative) for the underlying latent variable for each condition. The baseline scores can be used to infer the latent cutpoints for the model (projected onto the normal and cumulative distribution of the latent variable). These cutpoints can then be used to estimate the ordinal rating proportions for the experimental condition (illustrated by projecting the cutpoints derived from the baseline condition across the normal and cumulative distributions of the experiment latent variable).}
    \label{fig:step5}
\end{figure}

Figure \ref{fig:step5} shows this process. We infer the cutpoints from the baseline ratings, shift the mean of the normal distribution to the left (which implies an average decrease in the latent value) and see how it affects the distribution of scores. We offer a Shiny app in the supplementary material where you can set the cutpoints and observe how different effect sizes affect the distribution of scores.

So far, we described the intuition of the method generatively, that is, from the user's mind to the scores. In essence, CL(M)Ms do the inverse of this process: they take ordinal ratings and directly estimate the effects on the latent scale. Our example uses the cumulative proportion analogies to build an intuition. In practice, however, the cutpoints are not simply derived from the cumulative proportions of observed scores, but jointly estimated from the data along with the model coefficients. Next, we provide a more formal mathematical description of how it works and a practical implementation in R.

\subsection{Theory and Mathematics}
\label{sec:math_clmm}

As a predictive model, CL(M)Ms model the relationship between a dependent variable and predictor/independent variable(s). Let us extend the example used in section~\ref{sec:clmm_intuition} to illustrate this. Say we ask participants to use different versions of our system (a categorical independent variable represented as $X$) and rate their sense of spatial presence between 1 and 5, with 1 being `Very low' and 5 being `Very high' (an ordinal dependent variable represented as $Y$). We can represent this relationship as a causal graph, $X \rightarrow Y$, where the arrow pointing from $X$ (system version) to $Y$ (spatial presence response) indicates that $X$ has a causal effect on $Y$. In this scenario, the goal of predictive modelling would be to describe how changing the system version $X$ influences user responses about their sense of spatial presence $Y$, i.e., how $X$ predicts $Y$. However, the ordinal response, $Y$, is an \textit{imperfect measure} that the user generates about their sense of spatial presence in using a system version, based on some internal psychological measure and the available ordinal responses (elaborated in the example presented in figure~\ref{fig:step3}). For example, a user's true spatial presence experience may lie somewhere between `High' (scale value 4) and `Very High' (scale value 5), but the limited response options force a decision that imperfectly captures the user's true experience. Additionally, there is no guarantee that the internal psychological distance between two ordinal responses is evenly spaced. For instance, users may find it easy to change their response from 3 (`Moderate') to 4 (`High') --- small psychological distance --- but may reserve the highest response of 5 (`Very high') to systems that they perceive as exceptional at evoking a sense of spatial presence --- large psychological distance between 4 and 5 when compared to 3 and 4. This prevents meaningful analysis that relies on these distances to be consistent, such as statistical analysis methods that calculate averages or differences, and is a key cause of errors in analysing ordinal data using metric models~\cite{liddell2018analyzing,burkner2019ordinal}.  

\begin{figure}[tbp]
    \centering
    \includegraphics[width = 1.0\columnwidth]{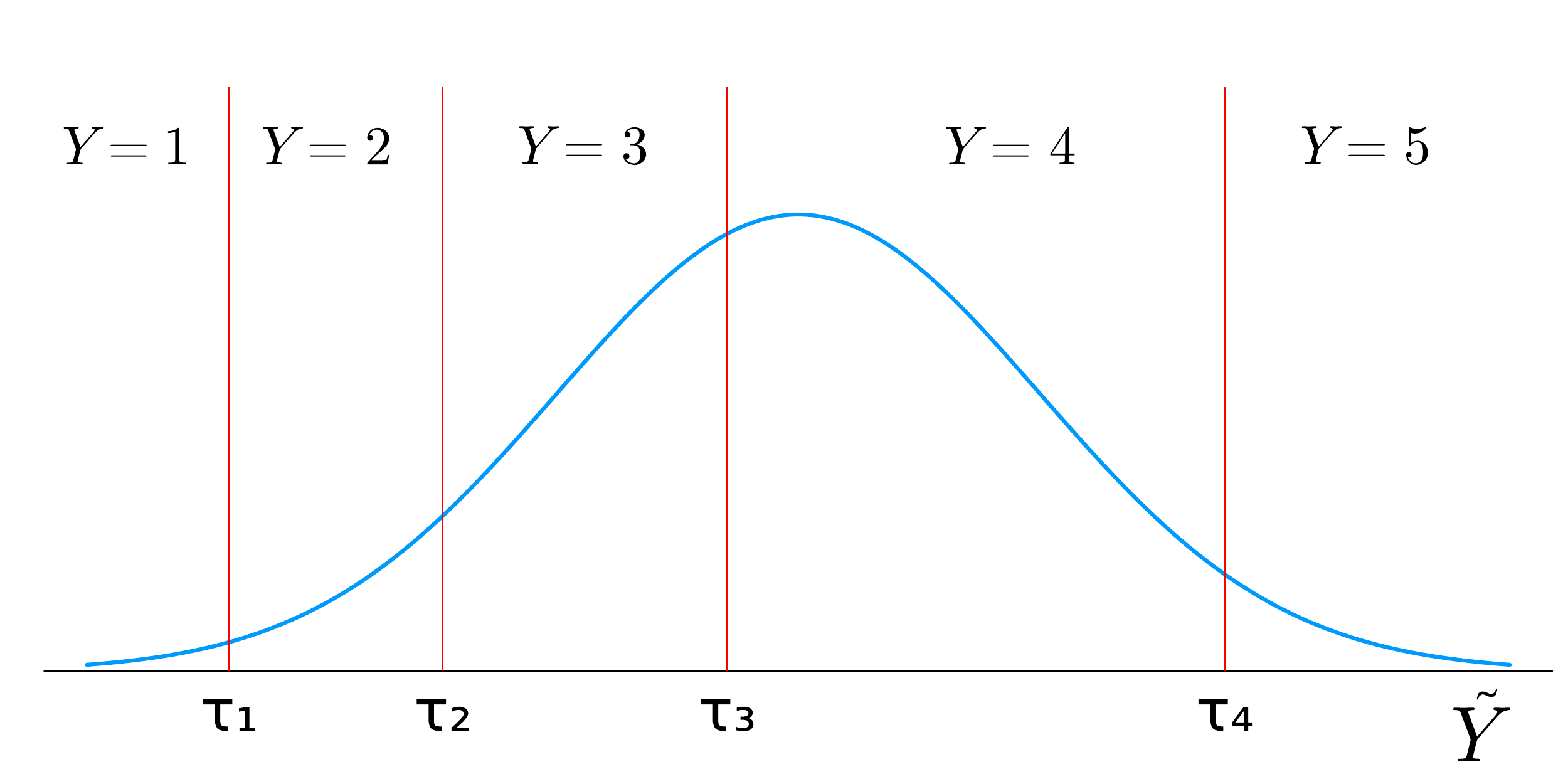}
    \caption{Figure illustrating the relationship between the internal continuous latent variable $\widetilde{Y}$ with the observable ordinal response $Y$ in our spatial presence example. Here, a user determines the system to elicit `Very low' spatial presence when their internal measure falls below the first cutpoint ($\Tau_1$), `Low' spatial presence, when their internal measure is between the first ($\Tau_1$) and second ($\Tau_2$) cutpoints, and so on.}
    \Description{Plot illustrating the relationship between the internal continuous latent variable (Y*) with the observable variable (Y) in our spatial presence example. Here, users respond with Y=1 (`Very low') when Y* is less than equal to cutpoint 1, Y=2 (`Low') when Y* is greater than cutpoint 1 and less that equal to cutpoint 2, and so on, with Y=5 (`Very high') when Y* is greater than cutpoint 4.}
    \label{fig:cutpoints_example}
\end{figure}

CL(M)Ms account for the internal psychological measure by assuming the presence of an underlying continuous latent (unobservable) variable (represented as $\widetilde{Y}$). This results in an updated causal graph, $X \rightarrow \widetilde{Y} \rightarrow Y$, which explicitly represents how the system version ($X$) affects an internal psychological measure of the user's sense of spatial presence ($\widetilde{Y}$), which is then categorized into an ordinal response ($Y$). Given this new causal structure, the relationship between the observable response, $Y$, and the latent variable, $\widetilde{Y}$, can then be defined using a set of \textbf{ordered thresholds/cutpoints, which determines the value that $Y$ categorizes to, when $\widetilde{Y}$ falls within certain cutpoints}. In our example, the 5 different responses (`Very low' to `Very high') will need 4 cutpoints (represented as $ \Tau = \{\Tau_1,\Tau_2,\Tau_3,\Tau_4\}$) to categorize all 5 available ordinal response, i.e, $Y=1$ (`Very low') when $\widetilde{Y} \le \Tau_1$, $Y=2$ (`Low') when $\Tau_1 < \widetilde{Y} \le \Tau_2$, and so on, with the highest value, $Y=5$ (`Very high') manifesting when $\Tau_4 < \widetilde{Y}$. This process is illustrated in Figure~\ref{fig:cutpoints_example} (and with a familiar Likert item example in figure~\ref{fig:step4}). In general, if we have $K$ different ordinal responses, we will need $K-1$ cutpoints. Formally:
\begin{equation}
    Y = k \text{ if } \Tau_{k-1} < \widetilde{Y} \le \Tau_k  \label{eq:1}
\end{equation}

This conceptualization of the latent \textit{continuous} variable, $\widetilde{Y}$, enables CL(M)Ms to mathematically express $Y$ in terms of $\widetilde{Y}$ and $\Tau$. CL(M)Ms can then model the continuous latent variable $\widetilde{Y}$ in relation to $X$ using linear regression. Collectively, these two steps enable CL(M)Ms to describe the relationships represented in the causal graph: $X \rightarrow \widetilde{Y} \rightarrow Y$.

To express $Y$ in terms of $\widetilde{Y}$ and $\Tau$, CL(M)Ms must first ensure that the observable ordinal data ($Y$) is appropriately treated as \textit{categorical}, avoiding over-representation of information (as with metric modelling approaches)~\cite{burkner2019ordinal}. However, treating ordinal data as categorical exposes only frequencies/probabilities as quantitative information, which are devoid of any ordering (under-representing available information as with most non-parametric approaches~\cite{burkner2019ordinal}). For example, the probability set \{0.2, 0.1, 0.7\} corresponding to an ordered dataset with levels \{A, B, C\} provides no information about the ordering of the data. To remedy this, CL(M)Ms make use of \textit{cumulative probabilities} in place of individual probabilities. This ensures that higher ordered categories are always associated with higher cumulative probabilities and vice-versa. Extending our example, the ordered dataset with levels \{A, B, C\} and probabilities \{0.2, 0.1, 0.7\}, will exhibit cumulative probabilities of \{0.2, 0.3, 1.0\}\footnote{Each value in this array is obtained by adding up all preceding values in the original array.}, preserving order. This also enables us to extract individual probabilities from cumulative probabilities using:
\begin{equation}
    P(Y=k) = P(Y \le k) - P(Y \le k-1) \label{eq:2}    
\end{equation}

Now, from~\eqref{eq:1}, we can infer that $Y \le k$ when $\widetilde{Y} \le \Tau_k$. Therefore, we can relate the cumulative probability $P(Y \le k)$ with the cumulative probability $P(\widetilde{Y} \le \Tau_k)$ which can be calculated using a cumulative distribution function over $\widetilde{Y}$, represented as $F(\Tau_k)$. Here, the function $F$ depends on the assumed distribution of $\widetilde{Y}$. Two commonly used distributions include the standard logistic distribution and the standard normal distributions, which have been referred to by the names ordered logit models and ordered probit models respectively~\cite{greene2010modeling}. The choice of assumed distribution influences the interpretation of the results but often results in similar parameter estimates and model fits~\cite{mccullagh1980regression}. For the purposes of this paper, we assume that $\widetilde{Y}$ follows a standard normal distribution. Details related to the different distributions and their corresponding cumulative distribution functions can be found in prior work~\cite{burkner2019ordinal,agresti2011categorical}. Given the function $F$, we can now express the relationship between $Y$ and $\widetilde{Y}$ as:
\begin{equation}
    P(Y \le k) = P(\widetilde{Y} \le \Tau_k) = F(\Tau_k) \label{eq:3}  
\end{equation}

From equation~\eqref{eq:2} and~\eqref{eq:3}, we can determine individual probabilities of ordinal responses based on the cumulative distribution function ($F$) and the cutpoints ($\Tau$):
\begin{equation}
    P(Y = k) = F(\Tau_k) - F(\Tau_{k-1}) \label{eq:4}  
\end{equation}

However, equation~\eqref{eq:4} does not explain the influence of system version ($X$) on spatial presence response ($Y$). To account for this, CL(M)Ms use linear regression to model the continuous latent variable $\widetilde{Y}$ with respect to $X$~\footnote{While we illustrate using a single independent variable, CL(M)Ms can handle multiple independent variables, where X would represent a matrix of column vectors.}:
\begin{equation}
    \widetilde{Y} = X^T\beta + \varepsilon \label{eq:5}
\end{equation}

Where the coefficient $\beta$ in the linear predictor ($X^T\beta$) describes the change in $\widetilde{Y}$ with respect to unit change in $X$, and $\varepsilon$ is the random error describing the assumed distribution of $\widetilde{Y}$ (standard normal in this example). To elaborate, let us assume that our spatial presence study example tests two system versions ($X_1 = \text{Version 1}$ and $X_2 = \text{Version 2}$), with the first version having slower screen frame rate than the second version. This can be represented in our linear model of $\widetilde{Y}$ as two binary (0 or 1) variables $X_1$ and $X_2$, indicating the presence or absence of that system version. Therefore, for this example, we can specify equation~\eqref{eq:5} as:
\begin{equation}
    \widetilde{Y} = X_1^T\beta_1 + X_2^T\beta_2 + \varepsilon \label{eq:6}
\end{equation}

Since a user can only experience one version at a time in our experiment that compares the two versions, either $X_1^T\beta_1$ or $X_2^T\beta_2$ will be equal to zero. Therefore, our linear predictor for $\widetilde{Y}$ can be described as:
\begin{align*}
    \widetilde{Y} =
    \begin{cases}
        X_1^T\beta_1 + \varepsilon, \text{ if } X = Version_1 \\
        X_2^T\beta_2 + \varepsilon, \text{ if } X = Version_2
    \end{cases}
\end{align*}

\begin{figure}[tbp]
    \centering
    \includegraphics[width = 1.0\columnwidth]{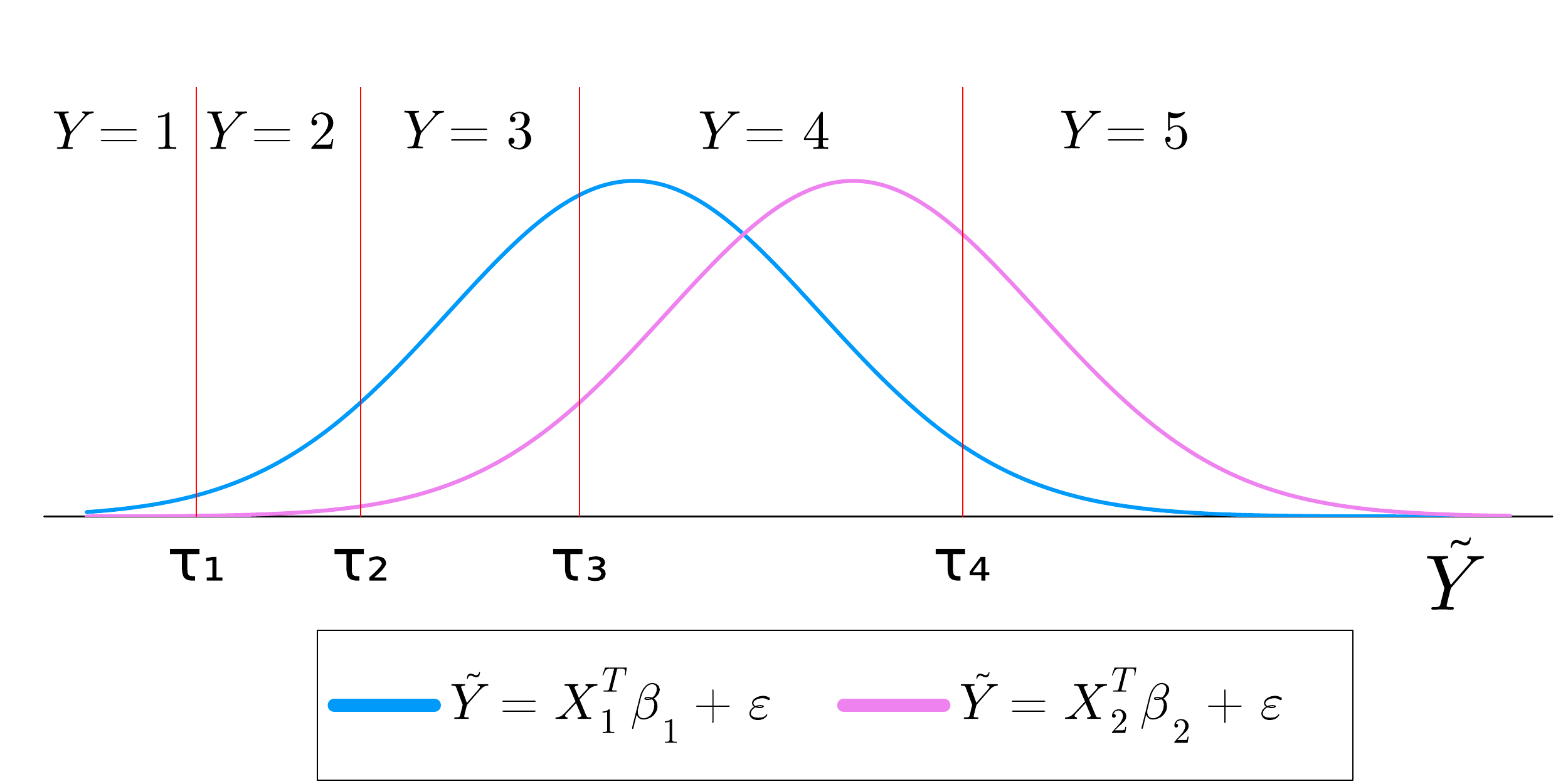}
    \caption{Figure showing example distributions of $\widetilde{Y}$ for $X = Version_1$ ($\widetilde{Y} =  X_1^T\beta_1 + \varepsilon$) and for $X = Version_2$ ($\widetilde{Y} =  X_2^T\beta_2 + \varepsilon$). The figure presents an example where users are more inclined to respond with a higher spatial presence rating to $Version_2$ compared to $Version_1$. This is depicted by the larger proportion of the curve (area under the curve) for $Version_2$ that falls between cutpoints ($\Tau$) that define higher ordinal responses, when compared to the curve for $Version_1$. For example, area under the curve for $ \Tau_4 < \widetilde{Y}$, defining the probability of a `Very high' user response, is larger for $Version_2$ than for $Version_1$} 
    \Description{Figure showing the distribution of the latent variable for two system versions (Y1* for Version 1 and Y2* for Version 2). The figure presents an example where users are more inclined to respond with a higher spatial presence rating Version 2 compared to Version 1. This is depicted by the larger proportion of the curve (area under the curve) for Version 2 that falls between cutpoints that define higher ordinal responses, when compared to the curve for Version 1.}
    \label{fig:linear_model_distribution_example}
\end{figure}

If Version 1 and 2 yield different effects on the latent spatial presence variable $\widetilde{Y}$ (i.e., $\beta_1  \neq \beta_2$), then $\widetilde{Y}$ will exhibit different \textit{distribution positions} for different system versions. As such, we can estimate the causal effect of switching from Version 1 to Version 2 by subtracting their respective means. Figure~\ref{fig:linear_model_distribution_example} illustrates the distribution of the latent spatial presence variable $\widetilde{Y}$ when users respond with a higher spatial presence rating for Version 2. 

Collectively, we now have equation~\eqref{eq:3} that describes the relationship between the user's response ($Y$) and their internal spatial presence measure ($\widetilde{Y}$), and equation~\eqref{eq:5} that models a linear relationship between the internal measure ($\widetilde{Y}$) and the system version ($X$). Therefore, we can model the probability of a user response ($Y$) being less than a specified value $k$ given the linear predictor ($X^T\beta$) of $\widetilde{Y}$ as: 
\begin{align}
    \begin{split}
        P(Y \le k|X^T\beta) = P(\widetilde{Y} \le \Tau_k|X^T\beta) = P(X^T\beta + \varepsilon \le \Tau_k) \\
    = P(\varepsilon \le \Tau_k - X^T\beta) = F(\Tau_k - X^T\beta) \label{eq:7}
    \end{split}
\end{align}

In practice, equation~\eqref{eq:7} is expressed in terms of the inverse cumulative distribution function ($F^{-1}$) which is the `link' function in generalised linear models to connect the linear predictor with the response value (see~\citet{agresti2011categorical} for a detailed description of these functions). Given~\eqref{eq:7}, we can update equation~\eqref{eq:4} to consider the linear predictor ($X^TB$) of $\widetilde{Y}$ to express the probability of $Y$ at individual responses as:
\begin{equation}
    P(Y=k|X^T\beta) = F(\Tau_k - X^T\beta) - F(\Tau_{k-1} - X^T\beta) \label{eq:8}
\end{equation}

This gives us the final equation that describes the relationship between the system version ($X$) and the observable spatial presence user rating ($Y$).

\subsection{Assumptions}
\label{sec:clmm_assumptions}
Like any statistical analysis method, CL(M)Ms impose a set of assumptions that should be carefully considered prior to their application. While we briefly mention these assumptions in sections~\ref{sec:clmm_intuition} \&~\ref{sec:math_clmm}, we elaborate on them here for further consideration. Additionally, tools to test for, and relax, some of the discussed assumptions are available with most software implementations for ordinal data analysis (see examples in R~\cite{christensen2018cumulative} and Stan~\cite{burkner2019ordinal}). If readers are not inclined towards making certain assumptions, and if these assumptions cannot be relaxed, they can refer to our table~\ref{tab:nonparametric_test_assumptions} for support in selecting alternative statistical tests to analyse their ordinal data.

\textbf{Latent Variable Assumption.} A fundamental assumption of CL(M)Ms relevant to HCI is that there exists an internal psychological measure (or latent variable) with an assumed distribution (standard normal in our `probit link' examples) that dictates, based on cutpoints, the ordinal response of a user to a specific survey question or Likert-item. The assumption of underlying latent variables is prominent and relevant in many fields, and particularly in psychology and the social sciences~\cite{bollen2002latent} where studies often focus on phenomena with directly unobservable causes. However, it is still worth considering whether such an assumption aligns with the researchers'/analysts' beliefs prior to the use of CL(M)Ms (or similar) for their analysis. 

\textbf{Latent Equal-Variance Assumption.} Heretofore, we have described CL(M)Ms using simplified examples that only illustrate changes in the location (distributional position) of the underlying latent variables $\widetilde{Y}$ (see figure~\ref{fig:step3},~\ref{fig:step5}, \&~\ref{fig:linear_model_distribution_example}). While standard CL(M)Ms assume constant variance of $\widetilde{Y}$, they can be extended to account for unequal variances by modelling the scale (standard deviation, $\sigma$) of $\widetilde{Y}$ within the model~\cite{christensen2018cumulative}. Practically, $\sigma$ can be incorporated through an additional regression factor in the model, often expressed as its inverse~\cite{burkner2019ordinal} ($\alpha = 1/\sigma$). Accordingly, equation~\eqref{eq:7} takes the form: $P(Y \le k|X^T\beta, \alpha) =  F(\alpha \times (\Tau_k - X^T\beta))$. Where, $\alpha = exp(X^T\beta_{\alpha})$ to ensure $\alpha > 0$, and $X^T\beta_{\alpha}$ is the linear predictor for $\alpha$. Modelling the scale as an additional regression factor enables the variance of $\widetilde{Y}$ to also be dependent on the variables $X$. More in-depth details can be found in~\citet{burkner2019ordinal}.

\textbf{Proportional Odds Assumption.} Another important assumption of CL(M)Ms is dubbed the \textit{proportional odds assumption}. This assumption refers to the unchanging coefficient estimates ($\beta$) across all cutpoint estimates ($\Tau = \{\Tau_1, \Tau_2, ..., \Tau_{k-1}\}$), i.e., the coefficients ($\beta$) remain the same $\forall\ k = \{1, 2, ..., K-1\}$ in $P(Y = k|X^T\beta)$ in equation~\eqref{eq:8}. \textit{Equivalently, this assumes that the cutpoints do not individually vary based on the independent variables ($X$)}. Effectively, this suggests that an intervention shifts user latent scores by the same amount across all cutpoints, and therefore has an equal effect on the likelihood of moving user responses from 1 to 2, 2 to 3, and so on. However, cases exist where an intervention's effect cannot be completely captured by a constant shift in the location of the original distribution, but also by how the distribution is `stretched' or `compressed' at different points. One way to test if this assumption holds is by checking for the equality of the scale $\sigma$ of the latent variable $\widetilde{Y}$ under different conditions, provided the model accounts for unequal variances, i.e., if the conditions only affect the location of the distribution, then the proportional odds assumption should hold. Statistical tests, such as the Brant test~\cite{brant1990assessing} can also be used to verify if this assumption holds. This assumption can also be partially relaxed~\cite{christensen2018cumulative, french2022regression} using the partial proportional odds model extension, which allows a subset of the independent variables to not assume proportional odds~\cite{peterson1990partial}. Here, an additional regression element ($P^T\gamma_k$) can then be incorporated into the model, where $P$ is a subset of the independent variables $X$, and $\gamma_k$ are the corresponding regression coefficients associated with $P$ for \textbf{the specific ordered category $k$}. The complete mathematical description around (partial) proportional odds is beyond the scope of this paper, and readers are referred to~\citet{peterson1990partial} for details.

While detailed mathematical knowledge of CL(M)Ms is not required for their use, a fundamental understanding of the latent model and its relationship to the ordinal data is essential for interpreting results and assessing assumptions. The following section moves from the theoretical aspects of CL(M)Ms to practical examples of analysing ordinal data using these models.

\section{Analysing Ordinal Data with CL(M)Ms in R}
\label{sec:analysing_using_clmms}

While historically less accessible~\cite{burkner2019ordinal, wobbrock2011the}, several software packages now exist that enable ordinal regression modelling in popular statistical tools used in HCI. These include the \texttt{statsmodel} package for Python~\cite{seabold2010statsmodels}, the \texttt{ologit} and \texttt{oprobit} functions for Stata~\footnote{\url{https://www.stata.com/features/overview/logistic-regression/}}, and the \texttt{repolr}~\cite{parsno2016repolr} and \texttt{brms} package in R~\cite{burkner2017brms}, among others. These packages provide different interfaces for fitting CL(M)Ms and may format model outputs differently. However, the underlying concepts (see Section~\ref{sec:clmm_theory}) behind CL(M)Ms remain unchanged. 

For the purposes of this paper, we demonstrate the practical use and interpretation of CL(M)Ms in HCI using R's \texttt{ordinal} package~\cite{christensen2023ordinal}. Specifically, we present two examples, using CLMs and CLMMs respectively, to reanalyses open-sourced data provided with previous CHI papers by~\citet{fitton2024watch} and~\citet{chen2024enhancing}. These papers were sampled in our initial search of the literature (see Section~\ref{sec:systematic_review}), but were excluded in our review during random sampling (Figure~\ref{fig:prisma-diagram}). Both examples make use of CL(M)Ms with a `probit' link to enable simpler result interpretation on the standard normal latent scale, as opposed to a log-odds interpretation when using the `logit' link. As mentioned in section~\ref{sec:math_clmm} and in prior work~\cite{mccullagh1980regression}, both link functions return similar results. Furthermore, for simplicity, our examples will assume equal variances and proportional odds. However, these assumptions can be relaxed when using the `ordinal' package by specifying the \textit{`scale'} and \textit{`nominal'} function variables respectively. Specific details for available functions in the \texttt{ordinal} package can be found in~\citet{christensen2018cumulative}. Finally, we present only a subset of the analysis plots in this section for illustrative purposes. Additional plots can be found in our Appendix~\ref{apx:additional_plots}. 

\paragraph{\textbf{Pre-processing and analysis replication:}} For each paper, we first load the available data into an R data frame and organise the data into a long-table format. This format eases use with R \textit{formulae} structure used in the `ordinal::clm()/clmm()' function. We then apply the analysis methods reported in the original paper to ensure that our pre-processing preserves data integrity and reproduces the findings reported. If test-statistics and p-values are provided in the paper, we ensure that these values are replicated in addition to alignment with significant/non-significant reports. For brevity, we do not reiterate the findings of the original analysis reported in the paper. Readers can refer to the original article or the analysis scripts provided in our supplementary material for additional details.

\subsection{CLM Example: ~\citet{fitton2024watch}}
\label{sec:clm_example}

\paragraph{Study Design \& Aim:} The paper by~\citet{fitton2024watch} explores how observing and performing a psychomotor task in virtual reality (VR) affects learning and retention. Four \textit{learning conditions} were evaluated: active learning (\textsc{Active}), observational learning with a self-avatar (\textsc{Self}), observational learning with a minimally similar avatar (\textsc{Minimal}), and observational learning with a dissimilar avatar (\textsc{Dissimilar}). These conditions were evaluated \textbf{between-subjects} using a puzzle assembly task. Immediately following the learning conditions in VR, participants were assessed in completing the puzzle in VR (immediate VR retention task) and in two real-world transfer assessments with 3D printed puzzle pieces (immediate near transfer task, immediate far transfer task). The near transfer task mirrored the learning condition in VR and presented the printed puzzle pieces as they appeared in VR (same colour coded pieces and initial relative piece pose), while the far transfer task differed from the learning condition by presenting non-coded puzzle pieces in different initial relative poses and included a second task of counting audio tones presented at random intervals. These tasks were performed by all participants, i.e., a \textit{within-subject} factor. The participants then returned after 10-14 days to complete the delayed VR retention task, delayed near transfer task, and the delayed far transfer task (\textit{within-subject}).

\paragraph{Outcome measures:} The paper presents several recorded outcome measures, including task completion times, number of pieces correctly assembled~\footnote{This was analysed using an ordinal regression model in the paper, but other ordinal data included in the paper were not.}, and questionnaire responses (Vividness of Imagery, Intrinsic Motivation Inventory, and Self-efficacy scores, among others). We focus on a subset of the ordinal responses analysed in the paper for illustrating the use of CLMs in HCI. Specifically, we re-analyse the:

\begin{enumerate}
    \item Usefulness of training (Perceived Usefulness), 1 Likert-type item.
    \item Simulated Task Load Index (SIM-TLX) survey, 9-item Likert-scale.
\end{enumerate}

This subset of measures were chosen as the raw, unprocessed data is provided, i.e., not averaged or weighted. Both measures were recorded \textbf{between-subjects} which enables analysis using CLMs.

\paragraph{Analysis used in the paper:} ~\citet{fitton2024watch} report a series of one-way ANOVAs to analyse the effect of learning condition on the `Perceived Usefulness' measure and on individual sub-scales of the SIM-TLX survey. The paper reports no significant differences between learning conditions on the `Perceived Usefulness' measure and across all SIM-TLX sub-scales. We were able to reproduce these results in our analysis by following the methods reported in the paper. 

\paragraph{\textbf{Modelling `Perceived Usefulness' using CLMs:}} We first organise data related to the `Perceived Usefulness' measure (\textsc{Usefulness}) and the corresponding learning condition (\textsc{Condition}) into an R data frame. The first 5 rows of the data frame is presented using the `head()' function:

\begin{shaded*}
{\small
\begin{verbatim}
head(usefulness_dataframe, 5)
  Condition  Usefulness
  <fct>        <ord>       
1 Dissimilar   5           
2 Self         4           
3 Dissimilar   2           
4 Active       5           
5 Self         5       
\end{verbatim}
}
\end{shaded*}

We can then fit a CLM with a probit link to the data, and display an overview of the model results by using the `clm()' function followed by the `summary()' function:

\begin{shaded*}
{\small
\begin{verbatim}
usefulness.clm <- clm(
    Usefulness ~ 1 + Condition, 
    data = usefulness_dataframe, 
    link = "probit")

summary(usefulness.clm)

formula: Usefulness ~ 1 + Condition
data:    usefulness_dataframe

link   threshold nobs logLik  AIC    niter max.grad cond.H 
probit flexible  102  -113.99 241.97 4(0)  1.47e-08 5.0e+01

Coefficients:
                Estimate Std. Error z value Pr(>|z|)
ConditionDissimilar -0.32161    0.32288  -0.996   0.319
ConditionSelf        0.04704    0.33048   0.142   0.887
ConditionMinimal    -0.49092    0.31466  -1.560   0.119

Threshold coefficients:
Estimate Std. Error z value
1|2  -2.5897     0.4409  -5.874
2|3  -1.8801     0.2965  -6.340
3|4  -1.1437     0.2519  -4.540
4|5  -0.2450     0.2358  -1.039
\end{verbatim}
}
\end{shaded*}

The `Coefficients' table and the `Threshold Coefficients' table present the primary results of model fitting. 

\paragraph{\textbf{Interpreting Results:}} The `Coefficients' table presents the estimated change ($\beta$ parameters) in the user's internal psychological measure of usefulness (latent variable, $\widetilde{Y}$) when the learning condition is changed from a reference condition. In this case, we set our reference condition as`\textsc{Condition = Active}', which, as described in Section~\ref{sec:clmm_theory}, has a standard normal distribution, $N(\mu = 0, \sigma^2 = 1)$. Therefore, the estimates can be interpreted as follows. The location (mean = $\mu$) of the user's internal measure of usefulness ($\widetilde{Y}$) distribution is shifted by $-0.322$ \textcolor{black}{($p=0.319$)} standard units (i.e. standard deviations) when changing the learning condition from \textsc{Active} to \textsc{Dissimilar}, $+0.047$ \textcolor{black}{($p=0.887$)} units from \textsc{Active} to \textsc{Self}, and $-0.491$ \textcolor{black}{($p=0.119$)} units from \textsc{Active} to \textsc{Minimal}. When assuming equal variances of the latent variables, this has a similar interpretation to Cohen's $d$ coefficient. Note that the summary presents p-values $\ge 0.5$ for each of the estimates ($\beta$), indicating that the estimates related to each of the non-reference learning conditions (\textsc{Dissimilar}, \textsc{Self}, \textsc{Minimal}) do not significantly differ from the reference condition (\textsc{Active}), i.e., these p-values are significance tests for the parameter estimates being zero (reference). 

The `Threshold Coefficients' table presents the cutpoint/threshold values (\Tau) relative to  $\widetilde{Y}$ at the reference condition (\textsc{Condition = Active}). These cutpoints determine the observable ordinal response measure of perceived usefulness ($Y$) as described by equation~\eqref{eq:1}. Specifically, user's will respond with a `Perceived Usefulness' ($Y$) measure of $Y = 1$ when their internal measure of usefulness ($\widetilde{Y}$) is $-2.590$ units below the standard normal mean (i.e., zero), $Y = 2$ when $ -2.590 < \widetilde{Y} \le -1.880$,  $Y = 3$ when $  -1.880 < \widetilde{Y} \le -1.144$,  $Y = 4$ when $-1.144 < \widetilde{Y} \le -0.245$, and $Y = 5$ when $-0.245 < \widetilde{Y}$. These estimates suggest that users are more likely to respond with $Y=5$ for \textsc{Condition = Active} given that the \textit{average} value of the internal psychological measure of usefulness ($0$ by assumption of standard normal distribution) is greater than the last cutpoint ($\Tau_4 = -0.245$). The frequency table (Table~\ref{tab:example_1_frequency_table}) for the responses based on the different learning conditions provides further support for this result --- indicating that the most common response for `Perceived Usefulness' in the \textsc{Active} condition is $Y=5$ (with 16/26 responses, or 61.5\% of all responses). 

\begin{table}[tbp]
    \caption{Frequency table of user responses (R) to the `Perceived Usefulness' survey in each learning condition (C) of {\protect\NoHyper~\citeauthor{fitton2024watch}'s}~\cite{fitton2024watch} paper.}
    \centering
        \begin{tabular}{c|ccccc}
            \toprule
            \diagbox{\textbf{C}}{\textbf{R}} & $Y=1$ & $Y=2$ & $Y=3$ & $Y=4$ & $Y=5$ \\
            \midrule
            Active & 0 & 1 & 3 & 6 & 16 \\
            Self & 0 & 0 & 3 & 7 & 15 \\
            Minimal & 1 & 0 & 4 & 12 & 9 \\
            Dissimilar & 0 & 3 & 3 & 6 & 13 \\
            \bottomrule
        \end{tabular}
        \label{tab:example_1_frequency_table}
\end{table}

To inspect the other learning conditions, we shift the location of the $\widetilde{Y}$ distribution from the standard normal (location at zero) by the coefficient estimate corresponding to the learning condition of interest, as presented in the `Coefficients table', i.e., $\widetilde{Y}_{Condition} \sim N(0 + Estimate_{Condition}, 1)$~\footnote{Note that the same results can be achieved by shifting the thresholds/cutpoints (\Tau) by the relevant estimate, instead of the location of $\widetilde{Y}$, as per equation~\eqref{eq:7}}. For instance, if we were interested in \textsc{Condition = Minimal}, we would shift the location of $\widetilde{Y}$ from the standard normal location ($\mu = 0$) by $-0.491$ units ($\mu = 0 - 0.491 = -0.491$). This suggests that for \textsc{Condition = Minimal}, user's will respond with $Y=1$ when $\widetilde{Y} - 0.491 \le -2.590$,  $Y = 2$ when $ -2.590 < \widetilde{Y} - 0.491 \le -1.880$,  $Y = 3$ when $  -1.880 < \widetilde{Y} - 0.491 \le -1.144$,  $Y = 4$ when $-1.144 < \widetilde{Y} - 0.491 \le -0.245$, and $Y = 5$ when $-0.245 < \widetilde{Y} - 0.491$. Note that this shifts the mean of $\widetilde{Y}$ for \textsc{Condition = Minimal} between cutpoints $3|4\ (\Tau_3$)  and $4|5\ (\Tau_4)$, suggesting that the most likely response, given a normal distribution, is $4$ (as $\Tau_3 < mean(\widetilde{Y}_{Minimal}) \le \Tau_4 => -1.144 < -0.491 \le -0.245$). This is also supported by Table~\ref{tab:example_1_frequency_table}, which indicates that the most frequent response for \textsc{Condition = Minimal} is $Y = 4$ with 12/26 of the total responses (46.2\%).

\paragraph{\textbf{Modelling SIM-TLX sub-scales using CLMs:}} As we have familiarised ourselves with CLM fitting and result interpretation in the previous sections, this example highlights differences in insights that may arise between statistical approaches traditionally used in HCI and CLMs. Specifically, we fit a CLM to each individual sub-scale of the SIM-TLX data provided by~\citet{fitton2024watch} to mirror their analysis using one-way ANOVAs. Contrary to their conclusion of no significant effects of learning condition on any of the SIM-TLX sub-scales, our analysis using CLMs suggests that parameter estimates for the sub-scales related to `Physical demands' and `Stress' shows significant divergence from zero for \textsc{Condition: Self}, i.e., significant differences between learning condition \textsc{Self} and the reference condition \textsc{Active}. We present the `Coefficients' table returned by the `summary()' functions for the CLMs fit to the `Physical demands' and `Stress' sub-scale data. The `Threshold Coefficients' table for the SIM-TLX measures are not presented here as details related to threshold/cutpoints interpretation is extensively discussed in the previous example. However, these details can be reproduced using our R scripts provided in the supplementary materials.

\begin{shaded*}
{\small
\begin{verbatim}
# SIM-TLX: Physical Demands
Coefficients:
                    Estimate Std. Error z value Pr(>|z|)  
ConditionDissimilar  -0.5196     0.2845  -1.826   0.0678 .
ConditionSelf        -0.6751     0.2884  -2.341   0.0192 *
ConditionMinimal     -0.1907     0.2799  -0.681   0.4958


# SIM-TLX: Stress
Coefficients:
                    Estimate Std. Error z value Pr(>|z|)  
ConditionDissimilar  0.09148    0.28315   0.323   0.7466  
ConditionSelf        0.59792    0.28613   2.090   0.0366 *
ConditionMinimal     0.43877    0.28198   1.556   0.1197 
\end{verbatim}
}
\end{shaded*}

\begin{figure}[t]
    \centering
    \begin{minipage}[t]{\columnwidth} %
        \centering
        \includegraphics[width=\columnwidth]{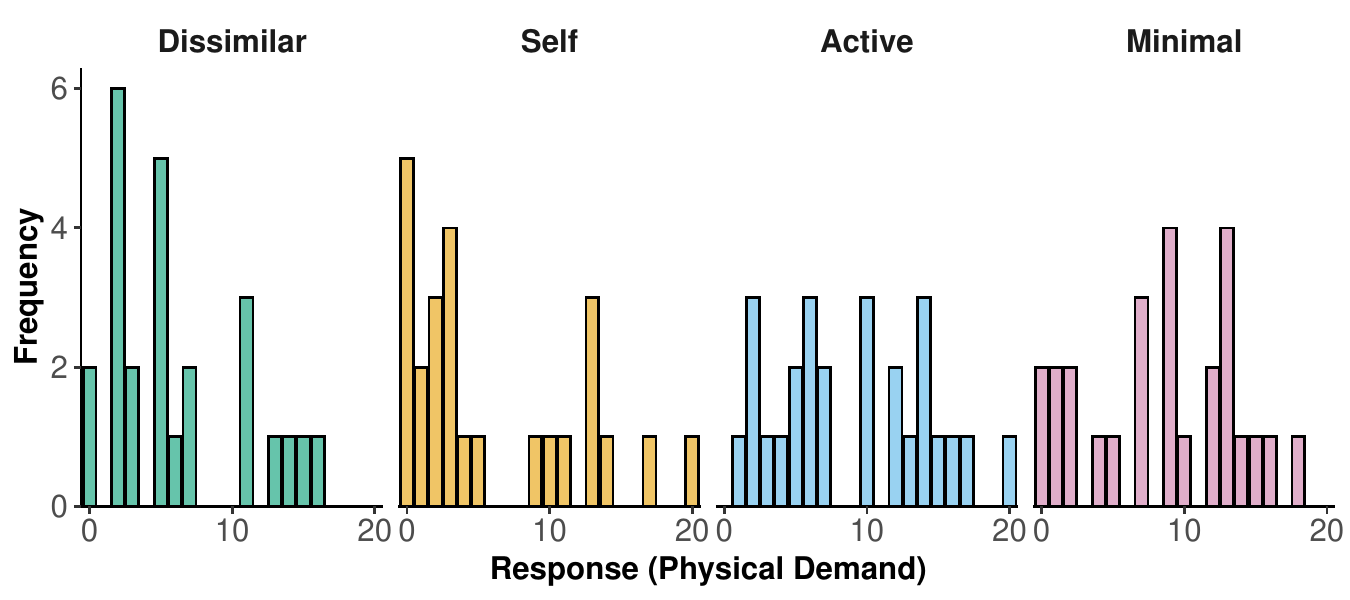}
    \end{minipage}
    \begin{minipage}[t]{\columnwidth} %
        \centering
        \includegraphics[width=\columnwidth]{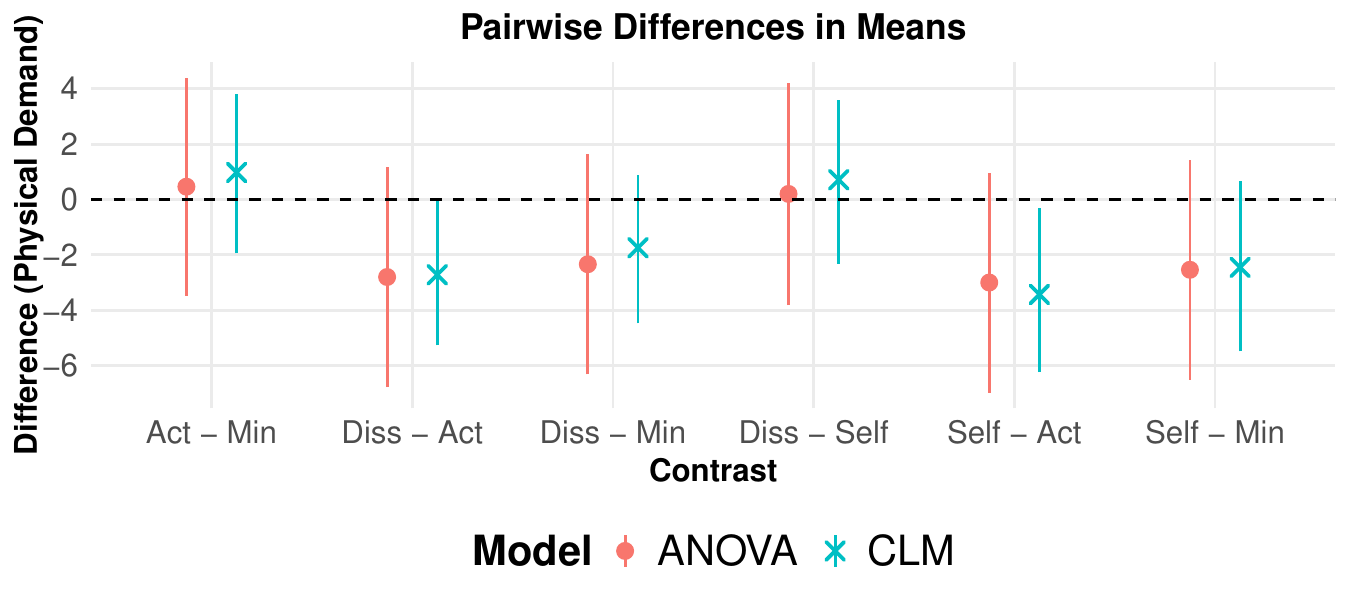}
    \end{minipage}
    \begin{minipage}[t]{\columnwidth} %
        \centering
        \includegraphics[width=\columnwidth]{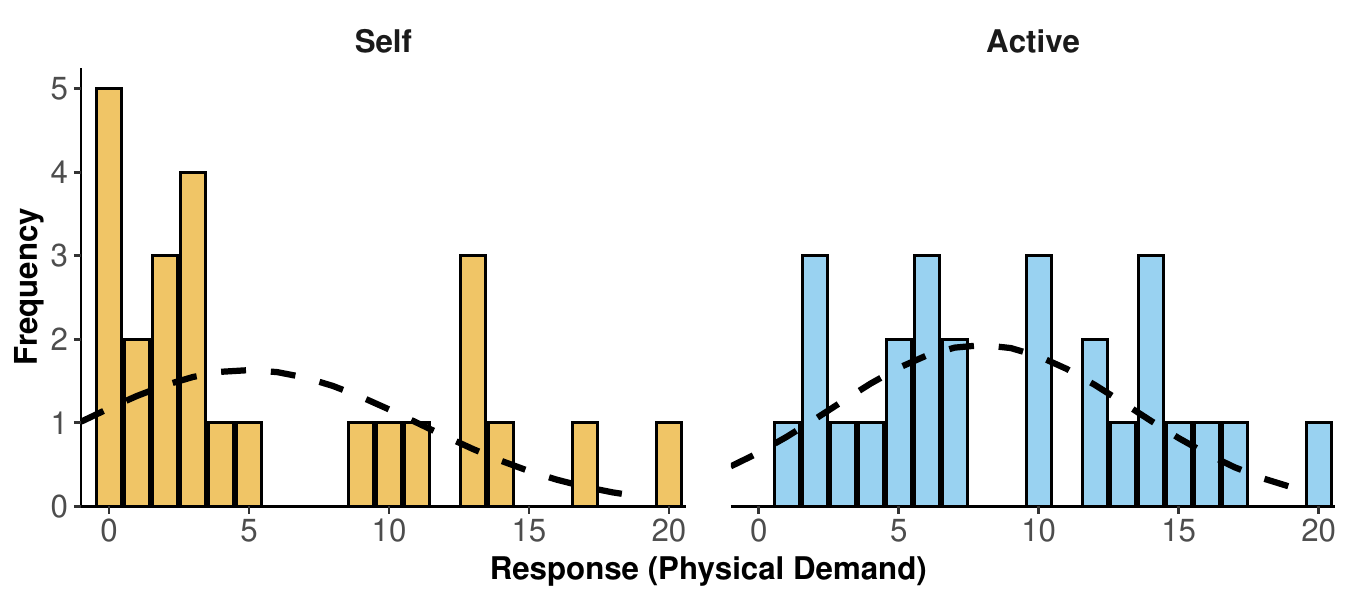}
    \end{minipage}
    \caption{Visualizations relevant to {\protect\NoHyper~\citeauthor{fitton2024watch}'s}~\cite{fitton2024watch} paper. (Top) Frequency of user responses on the `SIM-TLX: Physical Demand' subscale across learning conditions. (Middle) Confidence intervals for pairwise differences in mean responses between conditions under ANOVA and CLM models. (Bottom) Frequency plots for `Physical Demand' in the \textsc{Self} and \textsc{Active} conditions, which differed significantly under CLMs. Dashed curves show fitted linear (metric) model distributions based on condition means and standard deviations, highlighting how metric models can miss skewed responses. For example, \textsc{Self} responses are mostly minimal (0–3), but the metric model mean is inflated by a spike at 13.}
    \Description{Plots relevant to our re-analysis of the `SIM-TLX: Physical Demand' measure in Fitton et al.'s paper. The top figure shows a histogram of the `SIM-TLX: Physical Demand' ratings across the different conditions. The middle figure shows the Confidence Intervals for the pairwise differences of means between conditions under the ANOVA model used in the paper and our CLM model. The bottom figure shows histograms of the `Self' and `Active' conditions which differed significantly under the CLM model but not the ANOVA. The histograms are overlayed with a fitted linear (metric) model to illustrate how metric model fit the data.}
    \label{fig:physical_demand_plots}
\end{figure}

These findings suggest that the distribution for the user's internal measure of `Physical Demand' and `Stress' for the condition \textsc{Self} is significantly different from \textsc{Active}. These findings differ from the original analysis reported by~\citet{fitton2024watch}. To better illustrate these differences, figure~\ref{fig:physical_demand_plots} presents relevant plots related to the analysis of `Physical Demand'. Specifically, figure~\ref{fig:physical_demand_plots} presents the response frequency histograms, pairwise mean difference confidence intervals (CIs) under the ANOVA and CLM analysis, and a side-by-side comparison of the distributions (discrete and linear model fit) for `Physical Demand' responses under conditions where ANOVA and CLM analysis disagreed (\textsc{Self} and \textsc{Active}). Note that CIs for both models were obtained using the \texttt{emmeans} package in R~\cite{lenth2025emmeans}. Additionally, as CL(M)Ms report estimates on the latent scale, a bootstrapping approach~\cite{davison1997bootstrap} was employed to obtain CIs on the response scale for the CLM model to allow direct comparison with the ANOVA-based CIs in this example. Differences in CIs between the models suggest only a minor, but observable, disagreement. The side-by-side comparison, shown in the last subfigure, illustrates how metric model fits (based on the data's mean and standard deviations) can appear similar even when ordinal responses differ. For example, \textsc{Self} responses cluster at 0–3, but the metric model mean is skewed upward by a spike at 13, producing a distribution similar to condition \textsc{Active}~\footnote{This distributions did not appear normal. We confirmed this using a Shapiro-Wilk's test, and ran a Kruskal-Wallis non-parametric test between \textsc{Self} and \textsc{Active}, which also returned a non-significant difference.}. Similar examples with more prominent model disagreements, caused by ordinal responses at the extremes, can be found in {\protect\NoHyper~\citeauthor{liddell2018analyzing}'s}~\cite{liddell2018analyzing} paper.

Finally, if there is a need to predict estimates in relation to a reference condition other that the default (\textsc{Active}) condition, the `relevel()' function in the R `stats' package can be used to set the desired reference condition. While beyond the scope of this paper, post-hoc analysis of CL(M)Ms can be achieved by using appropriate methods, such as estimated marginal means~\cite{lenth2025emmeans} (\texttt{emmeans} R package), which would enable tests like multiple pairwise comparisons between the different conditions.

\subsection{CLMM Example: ~\citet{chen2024enhancing}}
\label{sec:clmm_example}

\paragraph{Study Design \& Aim:} {\protect\NoHyper~\citeauthor{chen2024enhancing}'s}~\cite{chen2024enhancing} paper explores the effects of different mechanism for adjusting display table height during follow-along home fitness instructional video viewing. Specifically, they investigate three conditions: \textsc{Fixed} --- where the display table height canot be adjusted, \textsc{Manual} --- where the user can manually adjust the display table height using buttons attached to the table, and \textsc{Automatic} --- where the height is automatically adjusted using a custom motion tracking system that aims to optimize viewing angles during fitness routines. The study employed a \textit{within-subject} design with randomized order for the conditions.

\paragraph{Outcome measures:} During the study, participant's head angle flexion was recorder, and an extended NASA-TLX survey~\cite{hart1988development} (scale 0 - 10) was administered after each condition. The extended TLX survey included two additional sub-scales related to `Degree of Support', and `Degree of Participation'. The paper reports the use of a weighting procedure on the NASA-TLX measures to capture individual differences in prioritizing workload dimensions. Weighting NASA-TLX measures is a common~\cite{hart1988development}, yet disputed practice~\cite{nygren1991psychometric,babaei2025should,virtanen2022weight}, and transforms the ordinal workload measures into interval scale which is incompatible for analysis using a CL(M)M. However, as~\citet{chen2024enhancing} also open-source the raw NASA-TLX measures, and because the \textit{purpose of this example is to illustrate the use of CLMMs for analysing data generated in HCI}, we focus solely on the raw ordinal workload measures. 

\paragraph{Analysis used in the paper:} The paper reports the use of a Friedman test with post-hoc Wilcoxon Signed Rank tests for pairwise comparisons between conditions of the \textit{weighted NASA-TLX measures}~\citet{chen2020examining}~\footnote{Additional analysis of the NASA-TLX measures with respect to participant related demographic factors were also conducted. We do not replicate these analyses as demographic data was not open-sourced.}. While we focus on the \textit{raw NASA-TLX measures}, we first run the analysis reported in the paper to ensure that we have correctly imported and organized the data in R. Our analysis using the reported Friedman test reproduced the exact p-values that were reported in the paper. However, we were unable to reproduce the pairwise significant differences reported in the paper using  post-hoc Wilcoxon Signed Rank tests. This may be because of a mismatch between our post-hoc test parameters and those used by~\citet{chen2024enhancing}, as the paper does not specify these details (such as the type of post-hoc correction used, if any). However, we conclude that we correctly imported the data into our R environment, as we reproduced the exact p-values reported for the Friedman test in the paper. We provide our analysis scripts in the supplementary material for transparency.

\paragraph{Modelling Raw NASA-TLX sub-scales using CLMMs:} The first 5 rows of our data frame containing the data from {\protect\NoHyper~\citeauthor{chen2024enhancing}'s}~\cite{chen2024enhancing} paper (named \textit{tlx\_df} in R) is presented using the `head()' function. It contains the raw NASA-TLX user score/response (\textit{score}), information regarding the experimental condition (\textit{condition}), information regarding the NASA-TLX sub-scale (\textit{tlx\_construct}), and the participant identifier (\textit{participant\_id}).

\begin{shaded*}
{\small
\begin{verbatim}
head(tlx_df,5)
  condition tlx_construct participant_id score
  <fct>     <chr>          <fct>         <dbl>
1 Fixed     Mental Demand   p1              5
2 Fixed     Mental Demand   p2              3
3 Fixed     Mental Demand   p3              2
4 Fixed     Mental Demand   p4              8
5 Fixed     Mental Demand   p5              2   
\end{verbatim}
}
\end{shaded*}

Fitting a CLMM to the data can be achieved by using the `ordinal' package's `clmm()'. Similar to the clm() function described in section~\ref{sec:clm_example}, we can use the `summary()' function to get an overview of the results. To illustrate the results returned by a CLMM, we present the output of fitting a CLMM to one example sub-scale (\textit{Physical Demand}) of the raw NASA-TLX measures provided in {\protect\NoHyper~\citeauthor{chen2024enhancing}'s}~\cite{chen2024enhancing} paper. Results for the remaining sub-scales can be found in our supplementary materials.  

\begin{shaded*}
{\footnotesize
\begin{verbatim}
# dataframe with only "Physical Demand" measures
pd_df <- tlx_df[
            tlx_df$tlx_construct == "Physical Demand",
            ]

physical_demand.clmm <- clmm(
    score ~ 1 + condition + (1|participant_id), 
    data = pd_df, 
    link = "probit")


summary(physical_demand.clmm)

formula: score ~ 1 + condition + (1 | participant_id)
data:    pd_df

 link   threshold nobs logLik  AIC    niter     max.grad cond.H 
 probit flexible  90   -161.75 343.50 706(2737) 3.18e-04 1.0e+02

Random effects:
 Groups         Name        Variance Std.Dev.
 participant_id (Intercept) 1.871    1.368   
Number of groups:  participant_id 30 

Coefficients:
                Estimate Std. Error z value Pr(>|z|)   
conditionFixed    0.4326     0.2777   1.558  0.11925   
conditionManual   0.7551     0.2844   2.655  0.00793 **
---
Signif. codes:  0 ‘***’ 0.001 ‘**’ 0.01 ‘*’ 0.05 ‘.’ 0.1 ‘ ’ 1

Threshold coefficients:
    Estimate Std. Error z value
2|3  -2.2670     0.4800  -4.723
3|4  -1.1828     0.3703  -3.194
4|5  -0.6183     0.3458  -1.788
5|6  -0.1894     0.3371  -0.562
6|7   0.3812     0.3390   1.124
7|8   1.3784     0.3651   3.775
8|9   2.5397     0.4396   5.777
\end{verbatim}
}
\end{shaded*}

Similar to the results returned by the `clm()' function, the `clmm()' summary returns the primary results of model fitting within the `Coefficients' and `Threshold Coefficients' tables. These can be interpreted in the same way as for the results return by the `clm()' summary, and is detailed in section~\ref{sec:clm_example}. In this example, the results suggest that the parameter estimate for the \textit{Manual} condition is significantly different from zero~\footnote{Greater than zero as the estimate is positive.}, i.e., the user's internal psychological measure of physical demand ($\widetilde{Y}$) for the \textit{Manual} condition is different from the reference condition (\textit{Automatic}) by an estimate that is significantly different from zero~\footnote{We refrain from making any comparisons with the analysis results presented in the original paper~\cite{chen2024enhancing} as our analysis focuses on the raw NASA-TLX measures as opposed to the weighted NASA-TLX measures used in the paper.}. The `Threshold Coefficients' table also suggest that the most likely response for the reference (\textit{Automatic}) condition is $Y = 6$ as the assumed location of the distribution ($\mu = 0$) falls between cutpoints 5|6 ($\Tau_5$) and 6|7 ($\Tau_6$). Note that insufficient data related to the lower responses (0 and 1) leads the `clmm()' summary to return a `Threshold Coefficients' table with only 7, out of 10, cutpoints, i.e., given 11 response categories (0 - 10) for the `Physical Demand' sub-scale used in the paper, there should be $(11 - 1) = 10$ cutpoints to partition the psychological measure ($\widetilde{Y}$) in 11 categories (see section~\ref{sec:clmm_theory}). 

Additionally, the CLMM model summary also returns a `Random effects' table which details group specific random intercept and slope variations (via the presented `Variance' and `Std.Dev'). In this case, our formula specifies that we want to model for the random intercept for each participant --- signified by the $(1|participant\_id)$ in our formula. The results suggest that participant's internal measure of physical demand varied by a standard deviation of $1.368$ units. This indicates a large variation between participants' psychological measures of `Physical demand', given the comparatively small values for cutpoint ($\Tau$) and parameter ($\beta$) estimates presented in the `Coefficients' and `Threshold Coefficients' tables. 

\subsection{Reporting CL(M)M Analysis Results}
\label{sec:reporting_clmm}
CLM/CLMM models report estimates (coefficients, std. error, etc.) relative to the latent variable scale, and is dependent on the `link' function used. Consequently, these values should be interpreted and reported considering these factors. As mentioned previously, both the `logit' and `probit' link functions return similar model fits but the returned estimates should be interpreted differently. We briefly describe `logit' model result interpretation and reporting, before detailing `probit' results and reporting. 

Ordinal `Logit' models return estimates as log-odds ratios. The coefficient estimate for a specific condition relative to the reference can be interpreted as the odds of that condition resulting in a higher response category than the reference condition~\cite{bruin2011newtest}. Let us remodel the `SIM-TLX:Physical Demand' measure presented in section~\ref{sec:clm_example} using a CLM with a `Logit' model. We forego presenting all estimates for brevity, but only present the significant coefficient for the \textsc{Self} condition relative to the reference (\textsc{Active}) for this illustration. The `logit' model returns a log-odds value of $-1.1951$, which when exponentiated (inverse log function) gives us the odds of observing higher ratings in the \textsc{Self} condition compared to the \textsc{Active} condition. This can be reported as: Users in the \textsc{Self} condition had an odds ratio of $exp(-1.1951) = 0.303$ compared to the \textsc{Active} condition, indicating they were significantly --- approximately 3.3 times ($1/0.303$) or $70\%$ ($1 - 0.303 \ or \ 100 - 30.3$) --- \textbf{less likely} to report higher physical demand than in the \textsc{Active} condition.

Ordinal `Probit' models return estimates on a standard normal scale for the assumed latent variable. This allows for direct interpretation of the effects on the latent scale, but prevents easy interpretation on the response or outcome scale. Let us take the same example `SIM-TLX:Physical Demand' used for illustration of the `logit' model. Here, the `probit' model returns a coefficient estimate of $-0.6751$ for the \textsc{Self} condition against the reference \textsc{Active} condition. This is interpreted as a \textit{significant reduction on the underlying latent score that determines the response of the user}, suggesting that users are less likely to report higher ratings for the \textsc{Self} condition compared to the \textsc{Active} condition. This can be reported as: Users in the \textsc{Self} condition were significantly less likely to report higher physical demand than those in the \textsc{Active} condition, with a decrease of $-0.6751$ on the latent score scale. The reliance on the latent scale makes `probit' model results easier to report for audiences familiar with latent models, foregoing the need to perform additional functions, such as exponentiation in case of `logit' models. However, this also implies that unfamiliar audiences would require appropriate prior introductions. As such, we recommend `probit' model use be accompanied by a brief acknowledgement of the underlying latent variable, along with appropriate references to more detailed literature on latent models (an example can be seen in the paper by~\citet{brailsford2024exploring}).

In addition to the above, reporting coefficient and threshold coefficient estimate means and standard errors for both ordinal `probit' and `logit' model results greatly help familiar readers in eliciting more detailed insights. We also recommend reporting assumptions, validity tests for assumptions, and if methods were employed to relax certain assumptions. Finally, we recommend accompanying these numbers and text with appropriate visualisations, such as for the frequency/proportions of the data, and confidence intervals of model estimates as shown inFigure~\ref{fig:physical_demand_plots} (top \& middle).  

\section{Discussion}

In this paper, we have shown the extent of divergent practices in analysing ordinal data within current HCI literature, discussed the associated concerns with the prevalent methods currently employed, and detailed the theory and application of CL(M)Ms as a more suitable method for ordinal data analysis with HCI related examples. Despite the stated advantages of CL(M)Ms over traditional statistical approaches for ordinal data analysis in HCI, certain limitations and assumptions should be considered.

Importantly, CL(M)Ms impose several assumptions that influence how results are interpreted, and these should be evaluated for validity and suitability. Take for example our re-analysis of the `Perceived Usefulness' measure provided by~\citet{fitton2024watch} in section~\ref{sec:clm_example}. Here, we used a CLM with the `probit' link function, and assumed proportional odds and equal variance. Use of the `probit' function forces interpretations on the latent scale. Specifically, effects of different conditions on `Perceived Usefulness' are estimated as shifts in a standard normal latent distribution, and cutpoints `divide' these latent distributions into proportions that correspond to the `Perceived Usefulness' ordinal categories. This is contrary to the use of the `logit' link  which enables interpretation as odds ratios (see section~\ref{sec:reporting_clmm}). Additionally, the proportional odds assumption enables us to inspect how estimates for conditions influence the latent distribution positions, without considering how different conditions affect the dividing cutpoints. This is evident in our elaboration, which highlights the shift in the latent distribution ($\widetilde{Y}$) and inspecting how the `static' cutpoints ($\Tau$) predict proportions of the ordinal response $Y$. However, such assumptions are often violated and should be tested for, and relaxed when necessary and possible (detailed in section~\ref{sec:clmm_assumptions}). This is also true for the equal variance of latent distribution assumption. But this can be relaxed using model extensions discussed in section~\ref{sec:clmm_assumptions}.

Both a strength and a limitation of CL(M)Ms lies in their \textit{design for analysing strictly ordinal data}. CL(M)Ms treat data as categorical with inherent ordering, enabling ordinal data analysis without requiring contiguous numerical labels for the ordered categories. As such, CL(M)Ms can be perfectly fit to data with ordinal responses that use non-numerical or non-contiguous labels (such as \{A,B,C\} or \{1,7, 91\}) as long as the order is defined. However, treating data as categorical also means that CL(M)Ms cannot handle ordinal data that has been transformed to interval or ratio scale, as such scales imply infinite categories, and therefore, infinite cutpoint estimates. This is evident by our inability to apply CLMMs to analyse the \textit{weighted} NASA-TLX measures in section~\ref{sec:clmm_example}, prompting us to instead focus on the \textit{raw} NASA-TLX measures provided by~\citet{chen2024enhancing}. Though possible, CL(M)Ms can also struggle to converge when applied to ordinal data with a large number of ordered categories (say a questionnaire with a slider response ranging from 0 to 100). In such cases where ordinal data needs to be transformed to metric data, or when the number of ordered categories is substantially large, it may be worth considering non-parametric alternatives that do not impose incompatible assumptions on ordinal data (see table~\ref{tab:nonparametric_test_assumptions}), or more specialized methods, such as beta regression models~\cite{geller2025beta}. While some prior work argue for the applicability of parametric methods with metric assumptions for analysing ordinal data under metric interpretations~\cite{carifio2007ten, carifio2008resolving, silan2025can, south2022effective}, the evidence for potential errors provided by~\citet{liddell2018analyzing} serves as a strong deterrent, and should be carefully considered when opting for such approaches.   

Finally, while we demonstrated Frequentist implementations of CL(M)Ms in this paper, these methods also have Bayesian counterparts. The \texttt{brms} package in R~\cite{burkner2017brms} uses a similar model specification syntax as the \texttt{ordinal} package. The biggest difference is in the requirement for the specification of the priors for the main effects (as in any Bayesian analysis) and for the cutpoints. An example of a default uninformative prior is offered by Kurz~\cite{kurz2022causal}, who uses normal distributions centred at the 1/(N-1) quantiles of the standard normal (where N is the number of possible scores), which implies a uniform prior for the distribution of scores. Examples of Bayesian analyses using ordinal regression models can be found in prior HCI work, including the use of CL(M)Ms with probit~\cite{wadinambiarachchi2024effects,brailsford2024exploring} and logit~\cite{hsu2025placebo} links.

\section{Conclusion}

As with most practices, it is pertinent to occasionally reflect on the methods that we use in HCI and update them where necessary. The case of ordinal data analysis within HCI is a prime example where such reflection can spotlight divisive opinions and approaches. Our review of the current HCI literature illustrates this well, highlighting not only disagreements in the use of parametric and non-parametric methods for ordinal data analysis between research articles, but also inconsistencies in the imposed data assumptions for ordinal data analysis between successive statistical procedures within the same study. Such inconsistencies hinder scientific progress by increasing the chances of propagating erroneous insights, introducing challenges to reproducible findings, and limiting comparisons between analytically distinct but related experiments. To address these issues, we propose the use of cumulative link (mixed) models, a form of ordinal regression modelling widely recommended for ordinal data analysis in disciplines such as psychology~\cite{liddell2018analyzing,burkner2019ordinal,christensen2018cumulative}. We explain the theory and mathematical foundations of CL(M)Ms, highlight its advantages over traditional methods used in HCI, and present practical examples of applying CL(M)Ms to published open-sourced ordinal data in HCI using the `ordinal' package in R~\cite{christensen2023ordinal}. Through these examples and the ensuing discussions, we provide HCI researchers with the necessary knowledge of when and how to apply CL(M)Ms to their data. We hope that our paper presents a convincing argument for the HCI community to adopt statistical approaches that are suitable for the intended data, and more broadly, to continue to reflect on, and update, our practices.

\bibliographystyle{ACM-Reference-Format}
\bibliography{bibliography}

@article{silan2025can,
  title={When can we treat Likert type data as interval?},
  author={Silan, Miguel Alejandro},
  year={2025}
}

@article{likert1932technique,
  title={A technique for the measurement of attitudes.},
  author={Likert, Rensis},
  journal={Archives of psychology},
  year={1932}
}

@inproceedings{dragicevic2019increasing,
  title={Increasing the transparency of research papers with explorable multiverse analyses},
  author={Dragicevic, Pierre and Jansen, Yvonne and Sarma, Abhraneel and Kay, Matthew and Chevalier, Fanny},
  booktitle={proceedings of the 2019 chi conference on human factors in computing systems},
  pages={1--15},
  year={2019}
}

@inproceedings{south2022effective,
  title={Effective use of Likert scales in visualization evaluations: A systematic review},
  author={South, Laura and Saffo, David and Vitek, Olga and Dunne, Cody and Borkin, Michelle A},
  booktitle={Computer Graphics Forum},
  volume={41},
  number={3},
  pages={43--55},
  year={2022},
  organization={Wiley Online Library}
}

@inproceedings{kaptein2010powerful,
  title={Powerful and consistent analysis of likert-type rating scales},
  author={Kaptein, Maurits Clemens and Nass, Clifford and Markopoulos, Panos},
  booktitle={Proceedings of the SIGCHI conference on human factors in computing systems},
  pages={2391--2394},
  year={2010}
}

@misc{carifio2008resolving,
  title={Resolving the 50-year debate around using and misusing Likert scales},
  author={Carifio, James and Perla, Rocco},
  journal={Medical education},
  volume={42},
  number={12},
  pages={1150--1152},
  year={2008},
  publisher={WILEY-BLACKWELL COMMERCE PLACE, 350 MAIN ST, MALDEN 02148, MA USA}
}

@article{carifio2007ten,
  title={Ten common misunderstandings, misconceptions, persistent myths and urban legends about Likert scales and Likert response formats and their antidotes},
  author={Carifio, James and Perla, Rocco J},
  journal={Journal of social sciences},
  volume={3},
  number={3},
  pages={106--116},
  year={2007},
  publisher={Science Publications}
}

@article{norman2010likert,
  title={Likert scales, levels of measurement and the “laws” of statistics},
  author={Norman, Geoff},
  journal={Advances in health sciences education},
  volume={15},
  pages={625--632},
  year={2010},
  publisher={Springer}
}

@article{jamieson2004likert,
  title={Likert scales: How to (ab) use them?},
  author={Jamieson, Susan},
  journal={Medical education},
  volume={38},
  number={12},
  pages={1217--1218},
  year={2004},
  publisher={Blackwell}
}

@article{calver2020anova,
  title={When ANOVA isn't ideal: Analyzing ordinal data from practical work in biology},
  author={Calver, Michael and Fletcher, Douglas},
  journal={The American Biology Teacher},
  volume={82},
  number={5},
  pages={289--294},
  year={2020},
  publisher={University of California Press}
}

@article{siegel1957nonparametric,
  title={Nonparametric statistics},
  author={Siegel, Sidney},
  journal={The American Statistician},
  volume={11},
  number={3},
  pages={13--19},
  year={1957},
  publisher={Taylor \& Francis}
}

@article{kornbrot1990rank,
  title={The rank difference test: A new and meaningful alternative to the Wilcoxon signed ranks test for ordinal data},
  author={Kornbrot, Diana Eugenie},
  journal={British Journal of Mathematical and Statistical Psychology},
  volume={43},
  number={2},
  pages={241--264},
  year={1990},
  publisher={Wiley Online Library}
}

@inproceedings{wobbrock2011the,
author = {Wobbrock, Jacob O. and Findlater, Leah and Gergle, Darren and Higgins, James J.},
title = {The aligned rank transform for nonparametric factorial analyses using only anova procedures},
year = {2011},
isbn = {9781450302289},
publisher = {Association for Computing Machinery},
address = {New York, NY, USA},
url = {https://doi.org/10.1145/1978942.1978963},
doi = {10.1145/1978942.1978963},
booktitle = {Proceedings of the SIGCHI Conference on Human Factors in Computing Systems},
pages = {143–146},
numpages = {4},
location = {Vancouver, BC, Canada},
series = {CHI '11}
}

@article{marcus1987meaningless,
  title={Meaningless statistics},
  author={Marcus-Roberts, Helen M and Roberts, Fred S},
  journal={Journal of Educational Statistics},
  volume={12},
  number={4},
  pages={383--394},
  year={1987},
  publisher={Sage Publications Sage CA: Los Angeles, CA}
}

@article{liddell2018analyzing,
  title={Analyzing ordinal data with metric models: What could possibly go wrong?},
  author={Liddell, Torrin M and Kruschke, John K},
  journal={Journal of Experimental Social Psychology},
  volume={79},
  pages={328--348},
  year={2018},
  publisher={Elsevier}
}

@article{mccullagh1980regression,
  title={Regression models for ordinal data},
  author={McCullagh, Peter},
  journal={Journal of the Royal Statistical Society: Series B (Methodological)},
  volume={42},
  number={2},
  pages={109--127},
  year={1980},
  publisher={Wiley Online Library}
}

@article{french2022regression,
  title={Regression models for ordinal outcomes},
  author={French, Benjamin and Shotwell, Matthew S},
  journal={Jama},
  volume={328},
  number={8},
  pages={772--773},
  year={2022},
  publisher={American Medical Association}
}

@article{peterson1990partial,
  title={Partial proportional odds models for ordinal response variables},
  author={Peterson, Bercedis and Harrell Jr, Frank E},
  journal={Journal of the Royal Statistical Society: Series C (Applied Statistics)},
  volume={39},
  number={2},
  pages={205--217},
  year={1990},
  publisher={Wiley Online Library}
}

@article{brant1990assessing,
  title={Assessing proportionality in the proportional odds model for ordinal logistic regression},
  author={Brant, Rollin},
  journal={Biometrics},
  pages={1171--1178},
  year={1990},
  publisher={JSTOR}
}

@article{burkner2019ordinal,
  title={Ordinal regression models in psychology: A tutorial},
  author={B{\"u}rkner, Paul-Christian and Vuorre, Matti},
  journal={Advances in Methods and Practices in Psychological Science},
  volume={2},
  number={1},
  pages={77--101},
  year={2019},
  publisher={Sage Publications Sage CA: Los Angeles, CA}
}

@article{gambarota2024ordinal,
  title={Ordinal regression models made easy: A tutorial on parameter interpretation, data simulation and power analysis},
  author={Gambarota, Filippo and Alto{\`e}, Gianmarco},
  journal={International Journal of Psychology},
  volume={59},
  number={6},
  pages={1263--1292},
  year={2024},
  publisher={Wiley Online Library}
}

@article{christensen2018cumulative,
  title={Cumulative link models for ordinal regression with the R package ordinal},
  author={Christensen, Rune Haubo B},
  journal={Submitted in J. Stat. Software},
  volume={35},
  pages={1--46},
  year={2018}
}

@misc{kurz2022causal,
  title = {Causal Inference with Ordinal Regresssion},
  howpublished = {\url{https://solomonkurz.netlify.app/blog/2023-05-21-causal-inference-with-ordinal-regression/}},
  note = {Accessed: 2025-08-28},
  author = {Solomon Kurz}
}

@online{katchova2020econometrics,
  author = {Ani Katchova},
  title = {Econometrics Academy},
  year = 2020,
  url = {https://sites.google.com/site/econometricsacademy/econometrics-models/ordered-probit-and-logit-models?authuser=0},
  urldate = {2025-06-19}
}

@article{williams2025cumulative,
  title={Cumulative Link Ordinal Outcome Neural Networks: An Evaluation of Current Methodology},
  author={Williams, Andre},
  journal={Journal of Data Analysis and Information Processing},
  volume={13},
  number={2},
  pages={182--198},
  year={2025},
  publisher={Scientific Research Publishing}
}

@article{chen2020examining,
  title={Examining injury severity in truck-involved collisions using a cumulative link mixed model},
  author={Chen, Mingyang and Chen, Peng and Gao, Xu and Yang, Chao},
  journal={Journal of Transport \& Health},
  volume={19},
  pages={100942},
  year={2020},
  publisher={Elsevier}
}

@article{schrum2023concerning,
  title={Concerning trends in likert scale usage in human-robot interaction: Towards improving best practices},
  author={Schrum, Mariah and Ghuy, Muyleng and Hedlund-Botti, Erin and Natarajan, Manisha and Johnson, Michael and Gombolay, Matthew},
  journal={ACM Transactions on Human-Robot Interaction},
  volume={12},
  number={3},
  pages={1--32},
  year={2023},
  publisher={ACM New York, NY}
}

@article{aguena2021clmm,
    author = {Aguena, M and Avestruz, C and Combet, C and Fu, S and Herbonnet, R and Malz, A I and Penna-Lima, M and Ricci, M and Vitenti, S D P and Baumont, L and Fan, H and Fong, M and Ho, M and Kirby, M and Payerne, C and Boutigny, D and Lee, B and Liu, B and McClintock, T and Miyatake, H and Sifón, C and von der Linden, A and Wu, H and Yoon, M and The LSST Dark Energy Science Collaboration },
    title = {CLMM: a LSST-DESC cluster weak lensing mass modeling library for cosmology},
    journal = {Monthly Notices of the Royal Astronomical Society},
    volume = {508},
    number = {4},
    pages = {6092-6110},
    year = {2021},
    month = {10},
    issn = {0035-8711},
    doi = {10.1093/mnras/stab2764},
    url = {https://doi.org/10.1093/mnras/stab2764},
    eprint = {https://academic.oup.com/mnras/article-pdf/508/4/6092/41074412/stab2764.pdf},
}

@book{greene2010modeling,
  title={Modeling ordered choices: A primer},
  author={Greene, William H and Hensher, David A},
  year={2010},
  publisher={Cambridge University Press}
}

@incollection{agresti2011categorical,
  title={Categorical data analysis},
  author={Agresti, Alan and Kateri, Maria},
  booktitle={International encyclopedia of statistical science},
  pages={206--208},
  year={2011},
  publisher={Springer}
}

@incollection{hart1988development,
  title={Development of NASA-TLX (Task Load Index): Results of empirical and theoretical research},
  author={Hart, Sandra G and Staveland, Lowell E},
  booktitle={Advances in psychology},
  volume={52},
  pages={139--183},
  year={1988},
  publisher={Elsevier}
}

@article{virtanen2022weight,
  title={Weight watchers: NASA-TLX weights revisited},
  author={Virtanen, Kai and Mansikka, Heikki and Kontio, Helmiina and Harris, Don},
  journal={TheoreTical issues in ergonomics science},
  volume={23},
  number={6},
  pages={725--748},
  year={2022},
  publisher={Taylor \& Francis}
}

@article{nygren1991psychometric,
  title={Psychometric properties of subjective workload measurement techniques: Implications for their use in the assessment of perceived mental workload},
  author={Nygren, Thomas E},
  journal={Human factors},
  volume={33},
  number={1},
  pages={17--33},
  year={1991},
  publisher={SAGE Publications Sage CA: Los Angeles, CA}
}

@article{babaei2025should,
  title={Should we use the NASA-TLX in HCI? A review of theoretical and methodological issues around Mental Workload Measurement},
  author={Babaei, Ebrahim and Dingler, Tilman and Tag, Benjamin and Velloso, Eduardo},
  journal={International Journal of Human-Computer Studies},
  pages={103515},
  year={2025},
  publisher={Elsevier}
}

@article{brooke1996sus,
  title={SUS-A quick and dirty usability scale},
  author={Brooke, John and others},
  journal={Usability evaluation in industry},
  volume={189},
  number={194},
  pages={4--7},
  year={1996},
  publisher={London, England.}
}

@misc{hayes2012process,
  title={PROCESS: A versatile computational tool for observed variable mediation, moderation, and conditional process modeling},
  author={Hayes, Andrew F},
  year={2012},
  publisher={University of Kansas, KS}
}

@Manual{lenth2025emmeans,
    title = {emmeans: Estimated Marginal Means, aka Least-Squares Means},
    author = {Russell V. Lenth},
    year = {2025},
    note = {R package version 1.10.7},
    url = {https://CRAN.R-project.org/package=emmeans},
  }

@book{davison1997bootstrap,
  title={Bootstrap methods and their application},
  author={Davison, Anthony Christopher and Hinkley, David Victor},
  number={1},
  year={1997},
  publisher={Cambridge university press}
}

@Manual{christensen2023ordinal,
    title = {ordinal---Regression Models for Ordinal Data},
    author = {Rune H. B. Christensen},
    year = {2023},
    note = {R package version 2023.12-4.1},
    url = {https://CRAN.R-project.org/package=ordinal},
  }

@article{burkner2017brms,
  title={brms: An R package for Bayesian multilevel models using Stan},
  author={B{\"u}rkner, Paul-Christian},
  journal={Journal of statistical software},
  volume={80},
  pages={1--28},
  year={2017}
}

@Manual{parsno2016repolr,
    title = {repolr: Repeated Measures Proportional Odds Logistic Regression},
    author = {Nick Parsons},
    year = {2016},
    note = {R package version 3.4},
    url = {https://CRAN.R-project.org/package=repolr},
  }

@inproceedings{seabold2010statsmodels,
  title={statsmodels: Econometric and statistical modeling with python},
  author={Seabold, Skipper and Perktold, Josef},
  booktitle={9th Python in Science Conference},
  year={2010},
}

@inproceedings{chen2024enhancing,
author = {Chen, Xinyu and Li, Yuqi and Chen, Jintao and Li, Jiabao and Wang, Chong and Tang, Pinyan},
title = {Enhancing Home Exercise Experiences with Video Motion-Tracking for Automatic Display Height Adjustment},
year = {2024},
isbn = {9798400703300},
publisher = {Association for Computing Machinery},
address = {New York, NY, USA},
url = {https://doi.org/10.1145/3613904.3642936},
doi = {10.1145/3613904.3642936},
booktitle = {Proceedings of the 2024 CHI Conference on Human Factors in Computing Systems},
articleno = {310},
numpages = {13},
location = {Honolulu, HI, USA},
series = {CHI '24}
}

@inproceedings{fitton2024watch,
author = {Fitton, Isabel Sophie and Dark, Elizabeth and Oliveira da Silva, Manoela Milena and Dalton, Jeremy and Proulx, Michael J and Clarke, Christopher and Lutteroth, Christof},
title = {Watch This! Observational Learning in VR Promotes Better Far Transfer than Active Learning for a Fine Psychomotor Task},
year = {2024},
isbn = {9798400703300},
publisher = {Association for Computing Machinery},
address = {New York, NY, USA},
url = {https://doi.org/10.1145/3613904.3642550},
doi = {10.1145/3613904.3642550},
booktitle = {Proceedings of the 2024 CHI Conference on Human Factors in Computing Systems},
articleno = {721},
numpages = {19},
location = {Honolulu, HI, USA},
series = {CHI '24}
}

@article{bollen2002latent,
  title={Latent variables in psychology and the social sciences},
  author={Bollen, Kenneth A},
  journal={Annual review of psychology},
  volume={53},
  number={1},
  pages={605--634},
  year={2002},
  publisher={Annual Reviews 4139 El Camino Way, PO Box 10139, Palo Alto, CA 94303-0139, USA}
}

@article{moher2009preferred,
  title={Preferred reporting items for systematic reviews and meta-analyses: the PRISMA statement},
  author={Moher, David and Liberati, Alessandro and Tetzlaff, Jennifer and Altman, Douglas G and PRISMA Group*},
  journal={Annals of internal medicine},
  volume={151},
  number={4},
  pages={264--269},
  year={2009},
  publisher={American College of Physicians}
}

@article{dunn1964multiple,
  title={Multiple comparisons using rank sums},
  author={Dunn, Olive Jean},
  journal={Technometrics},
  volume={6},
  number={3},
  pages={241--252},
  year={1964},
  publisher={Taylor \& Francis}
}

@article{dinno2015nonparametric,
  title={Nonparametric pairwise multiple comparisons in independent groups using Dunn's test},
  author={Dinno, Alexis},
  journal={The Stata Journal},
  volume={15},
  number={1},
  pages={292--300},
  year={2015},
  publisher={SAGE Publications Sage CA: Los Angeles, CA}
}

@techreport{conover1979multiple,
  title={Multiple-comparisons procedures. Informal report},
  author={Conover, WJ and Iman, RL},
  year={1979},
  institution={Los Alamos National Lab.(LANL), Los Alamos, NM (United States)}
}

@book{conover1999practical,
  title={Practical nonparametric statistics},
  author={Conover, William Jay},
  year={1999},
  publisher={john wiley \& sons}
}

@Manual{pohlert2024the,
    title = {PMCMRplus: Calculate Pairwise Multiple Comparisons of Mean
      Rank Sums Extended},
    author = {Thorsten Pohlert},
    year = {2024},
    note = {R package version 1.9.12},
    url = {https://CRAN.R-project.org/package=PMCMRplus},
  }

@inproceedings{elkin2021aligned,
  title={An aligned rank transform procedure for multifactor contrast tests},
  author={Elkin, Lisa A and Kay, Matthew and Higgins, James J and Wobbrock, Jacob O},
  booktitle={The 34th annual ACM symposium on user interface software and technology},
  pages={754--768},
  year={2021}
}

@article{wilcoxon1945individual,
  title={Individual comparisons by ranking methods},
  author={Wilcoxon, Frank},
  journal={Biometrics bulletin},
  volume={1},
  number={6},
  pages={80--83},
  year={1945},
  publisher={JSTOR}
}

@article{pearson1900x,
  title={X. On the criterion that a given system of deviations from the probable in the case of a correlated system of variables is such that it can be reasonably supposed to have arisen from random sampling},
  author={Pearson, Karl},
  journal={The London, Edinburgh, and Dublin Philosophical Magazine and Journal of Science},
  volume={50},
  number={302},
  pages={157--175},
  year={1900},
  publisher={Taylor \& Francis}
}

@book{nemenyi1963distribution,
  title={Distribution-free multiple comparisons.},
  author={Nemenyi, Peter Bjorn},
  year={1963},
  publisher={Princeton University}
}

@article{games1976pairwise,
  title={Pairwise multiple comparison procedures with unequal n’s and/or variances: a Monte Carlo study},
  author={Games, Paul A and Howell, John F},
  journal={Journal of Educational Statistics},
  volume={1},
  number={2},
  pages={113--125},
  year={1976},
  publisher={Sage Publications Sage CA: Thousand Oaks, CA}
}

@article{shingala2015comparison,
  title={Comparison of post hoc tests for unequal variance},
  author={Shingala, Mital C and Rajyaguru, Arti},
  journal={International Journal of New Technologies in Science and Engineering},
  volume={2},
  number={5},
  pages={22--33},
  year={2015}
}

@article{berger2014kolmogorov,
  title={Kolmogorov--smirnov test: Overview},
  author={Berger, Vance W and Zhou, YanYan},
  journal={Wiley statsref: Statistics reference online},
  year={2014},
  publisher={Wiley Online Library}
}

@article{kruskal1952use,
  title={Use of ranks in one-criterion variance analysis},
  author={Kruskal, William H and Wallis, W Allen},
  journal={Journal of the American statistical Association},
  volume={47},
  number={260},
  pages={583--621},
  year={1952},
  publisher={Taylor \& Francis}
}

@article{friedman1937use,
  title={The use of ranks to avoid the assumption of normality implicit in the analysis of variance},
  author={Friedman, Milton},
  journal={Journal of the american statistical association},
  volume={32},
  number={200},
  pages={675--701},
  year={1937},
  publisher={Taylor \& Francis}
}

@inproceedings{wadinambiarachchi2024effects,
  title={The effects of generative ai on design fixation and divergent thinking},
  author={Wadinambiarachchi, Samangi and Kelly, Ryan M and Pareek, Saumya and Zhou, Qiushi and Velloso, Eduardo},
  booktitle={Proceedings of the 2024 CHI Conference on Human Factors in Computing Systems},
  pages={1--18},
  year={2024}
}

@inproceedings{brailsford2024exploring,
  title={Exploring the association between moral foundations and judgements of AI behaviour},
  author={Brailsford, Joe and Vetere, Frank and Velloso, Eduardo},
  booktitle={Proceedings of the 2024 CHI Conference on Human Factors in Computing Systems},
  pages={1--15},
  year={2024}
}

@inproceedings{hsu2025placebo,
  title={Placebo Effect of Control Settings in Feeds Are Not Always Strong},
  author={Hsu, Silas and Koshy, Vinay and Vaccaro, Kristen and Sandvig, Christian and Karahalios, Karrie},
  booktitle={Proceedings of the 2025 CHI Conference on Human Factors in Computing Systems},
  pages={1--16},
  year={2025}
}

@misc{geller2025beta,
 title={A Beta Way: A Tutorial For Using Beta Regression in Psychological Research},
 url={osf.io/preprints/psyarxiv/d6v5t_v2},
 DOI={10.31234/osf.io/d6v5t_v2},
 publisher={PsyArXiv},
 author={Geller, Jason and Vuorre, Matti and Parlett, Chelsea and Kubinec, Robert},
 year={2025},
 month={Aug}
}

@ONLINE{bruin2011newtest,
  author = {Bruin, J.},
  title = {newtest: command to compute new test {@ONLINE}},
  month = FEB,
  year = {2011},
  url = {https://stats.oarc.ucla.edu/other/mult-pkg/faq/ologit/}
}

\appendix

\clearpage

\onecolumn

\section{Appendix}
\subsection{Additional Plots}
\label{apx:additional_plots}

\begin{figure}[h]
    \centering
    \begin{minipage}[t]{0.7\columnwidth}
        \centering
        \includegraphics[width=\columnwidth]{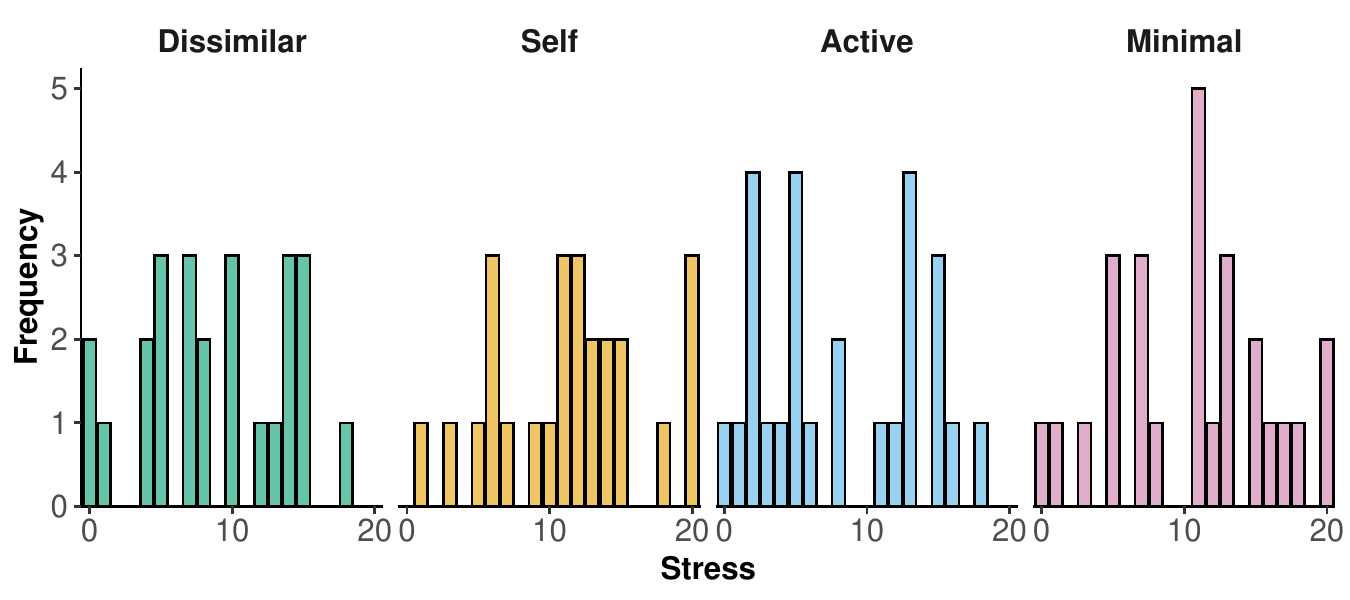}
    \end{minipage}
    \begin{minipage}[t]{0.7\columnwidth}
        \centering
        \includegraphics[width=\columnwidth]{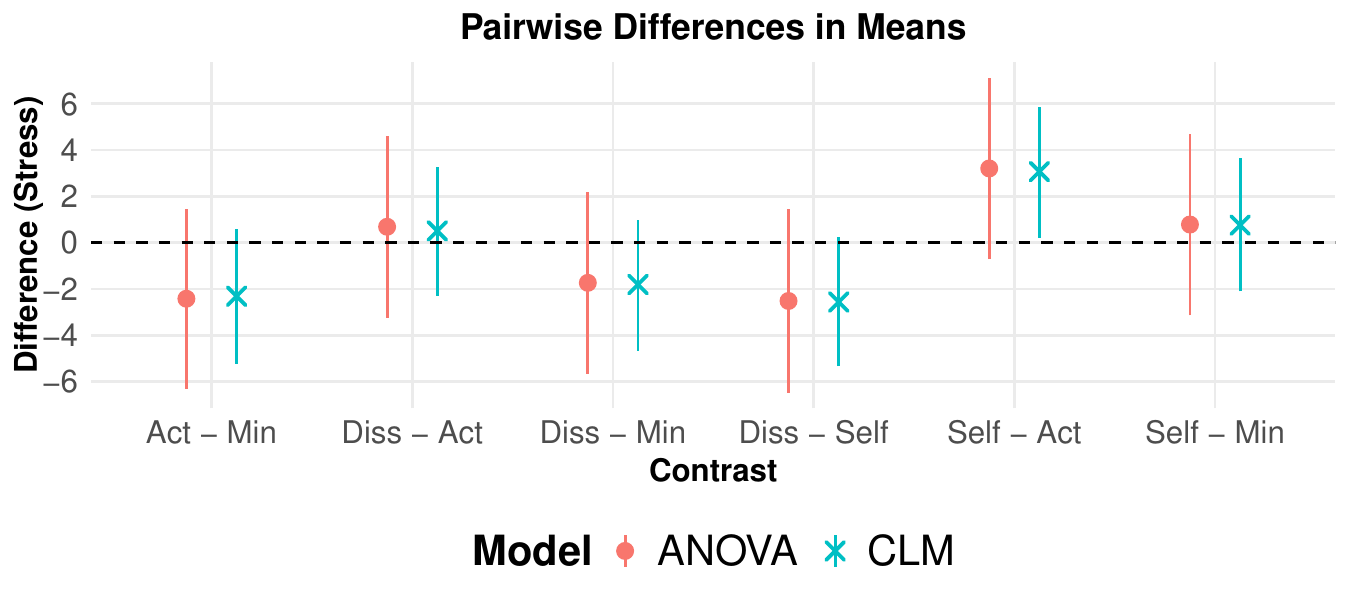}
    \end{minipage}
    \begin{minipage}[t]{0.7\columnwidth}
        \centering
        \includegraphics[width=\columnwidth]{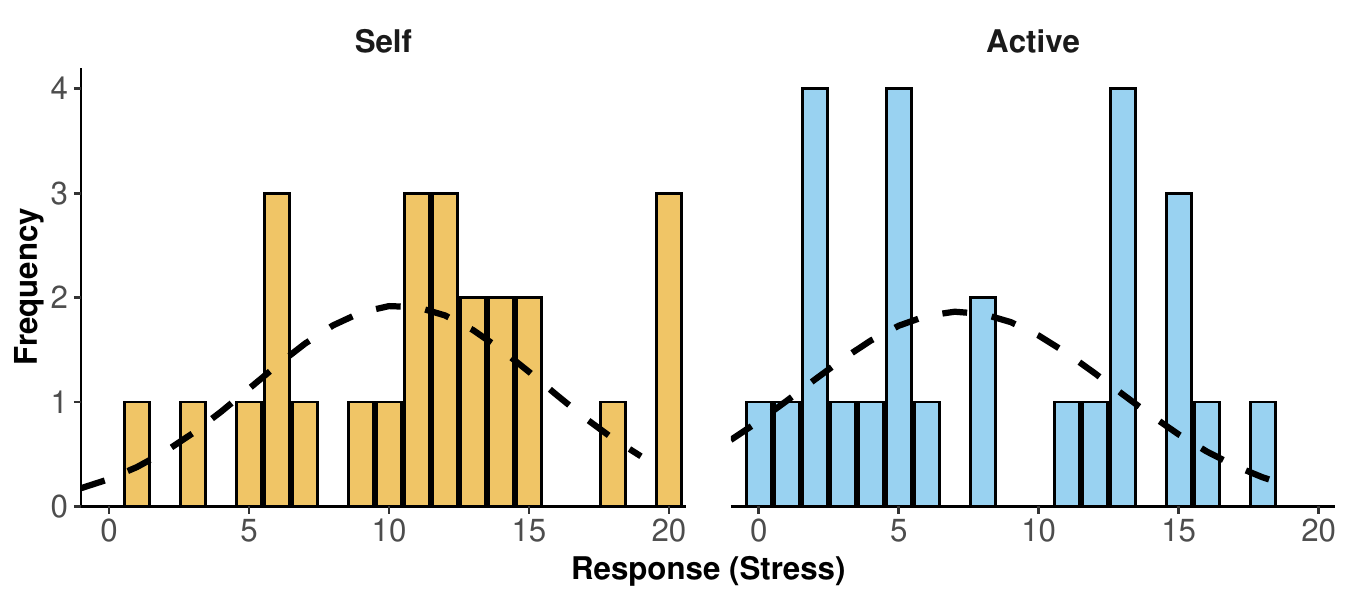}
    \end{minipage}
    \caption{Visualizations relevant to {\protect\NoHyper~\citeauthor{fitton2024watch}'s}~\cite{fitton2024watch} paper. (Top) Frequency of user responses on the `SIM-TLX: Stress' sub-scale across learning conditions. (Middle) Confidence intervals for pairwise differences in mean responses between conditions under ANOVA and CLM models. (Bottom) Frequency plots for `Stress' in the \textsc{Self} and \textsc{Active} conditions, which differed significantly under CLMs. Dashed curves show fitted linear (metric) model distributions based on condition means and standard deviations.}
    \Description{Plots relevant to our re-analysis of the `SIM-TLX:Stress' measure in Fitton et al.'s paper. The top figure shows a histogram of the `SIM-TLX: Stress' ratings across the different conditions. The middle figure shows the Confidence Intervals for the pairwise differences of means between conditions under the ANOVA model used in the paper and our CLM model. The bottom figure shows histograms of the `Self' and `Active' conditions which differed significantly under the CLM model but not the ANOVA. The histograms are overlayed with a fitted linear (metric) model to illustrate how metric model fit the data.}
    \label{fig:stress_plots}
\end{figure}

\begin{figure}[h]
    \centering
    \begin{minipage}[t]{0.7\columnwidth}
        \centering
        \includegraphics[width=\columnwidth]{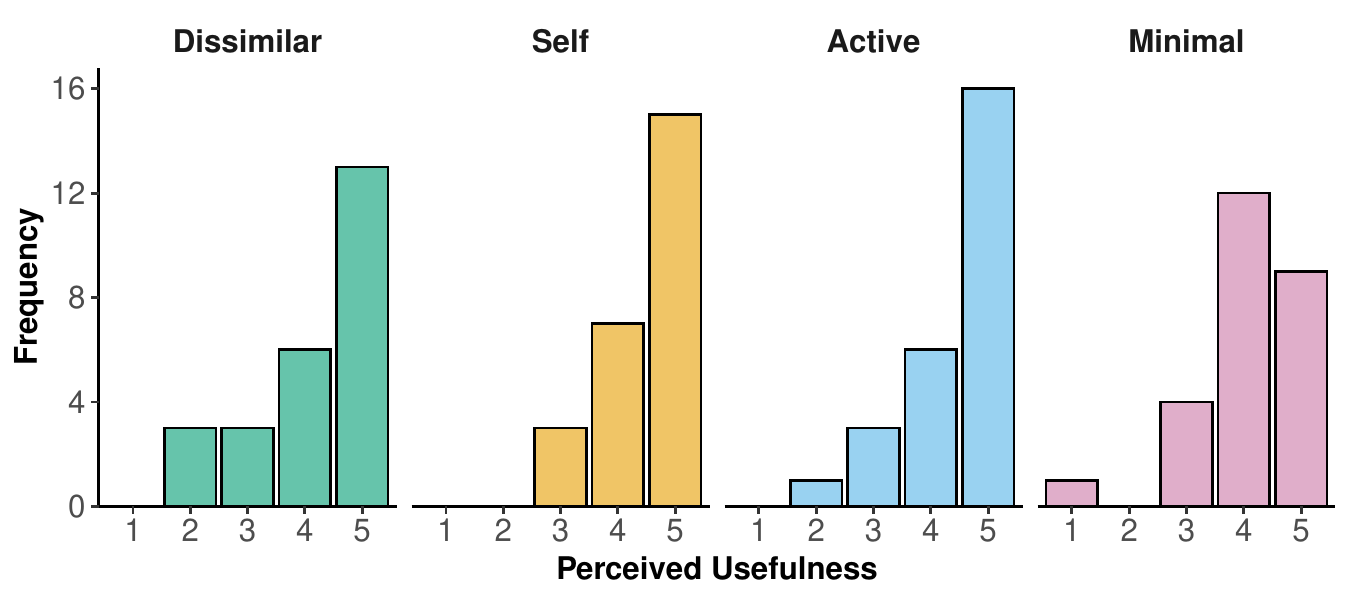}
    \end{minipage}
    \begin{minipage}[t]{0.7\columnwidth}
        \centering
        \includegraphics[width=\columnwidth]{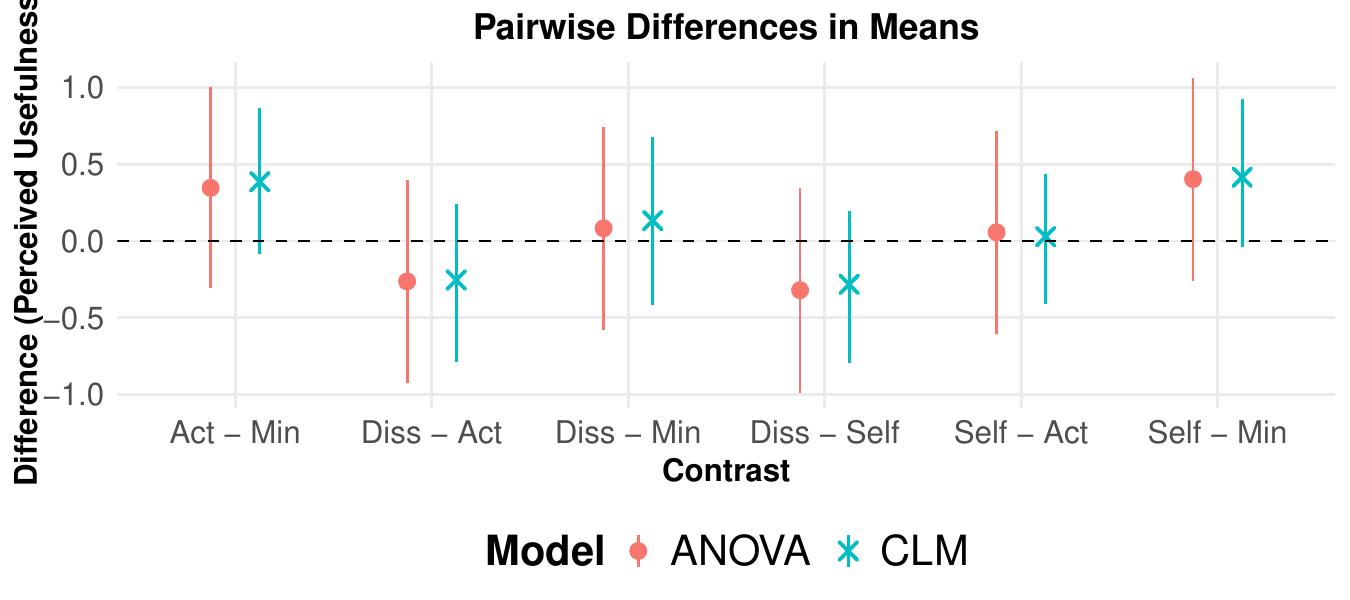}
    \end{minipage}
    \caption{Visualizations relevant to {\protect\NoHyper~\citeauthor{fitton2024watch}'s}~\cite{fitton2024watch} paper. (Top) Frequency of user responses to the `Perceived Usefulness' measure across learning conditions. (Middle) Confidence intervals for pairwise differences in mean responses between conditions under ANOVA and CLM models. No disagreements were found between the ANOVA and CLM model analyses.}
    \Description{Plots relevant to our re-analysis of the `Perceived Usefulness' measure in Fitton et al.'s paper. The top figure shows a histogram of the `Perceived Usefulness' ratings across the different conditions. The middle figure shows the Confidence Intervals for the pairwise differences of means between conditions under the ANOVA model used in the paper and our CLM model. No disagreements were observed between the ANOVA and CLM models.}
    \label{fig:usefulness_plots}
\end{figure}

\begin{figure}
    \centering
    \includegraphics[width=0.7\columnwidth]{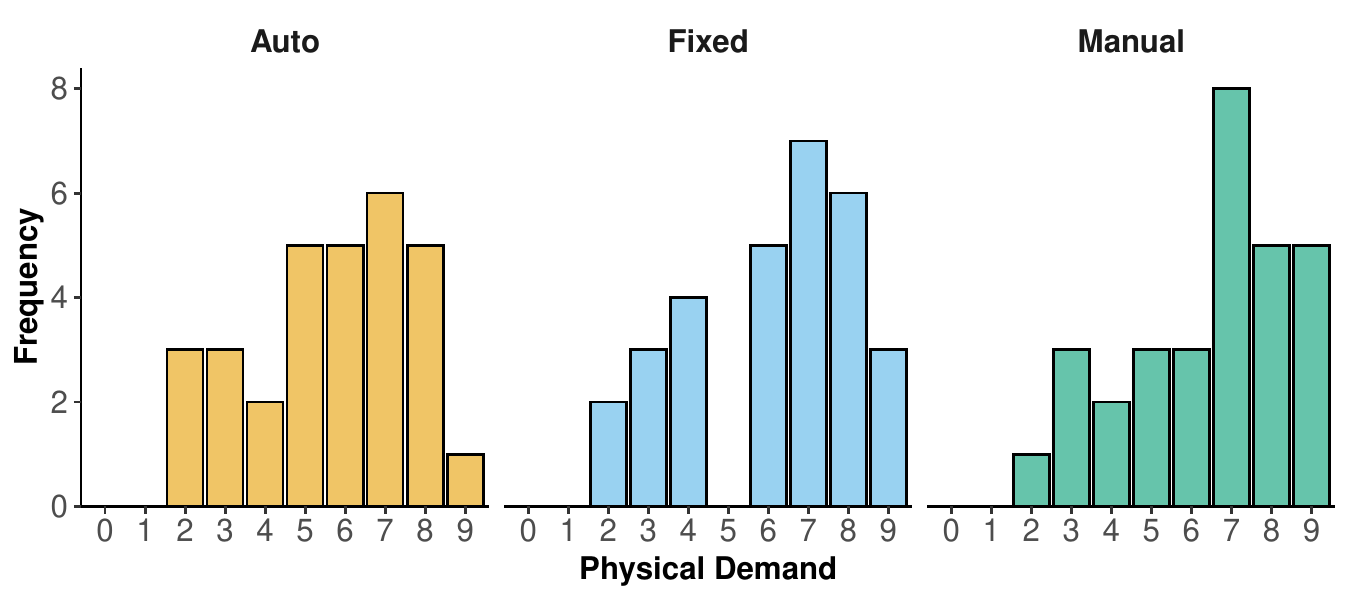}
    \caption{Visualizations relevant to {\protect\NoHyper~\citeauthor{chen2024enhancing}'s}~\cite{chen2024enhancing} paper. (Top) Frequency of user responses on the `NASA-TLX: Physical Demand' sub-scale across conditions. Comparative plots such as mean difference confidence intervals between modelling approaches are not presented as {\protect\NoHyper~\citeauthor{chen2024enhancing}'s}~\cite{chen2024enhancing} paper uses weighted NASA-TLX scores which cannot be analysed using an ordinal regression model like CLMM. This prevented us from plotting any comparative plots between statistical analysis methods used in the paper and our re-analysis using a CLMM.}
    \Description{Plots relevant to our re-analysis of the `NASA-TLX: Physical Demand' measure in Chen et al.'s paper. The figure shows a histogram of the unweighted `NASA-TLX: Physical Demand' ratings across the different conditions. As Chen et al's paper uses the weighted NASA-TLX scores which cannot be analysed using CL(M)Ms, no comparison plots are presented.}
    \label{fig:placeholder}
\end{figure}

\clearpage

\begin{figure*}[h]
    \centering
    \includegraphics[width = 1.0\textwidth]{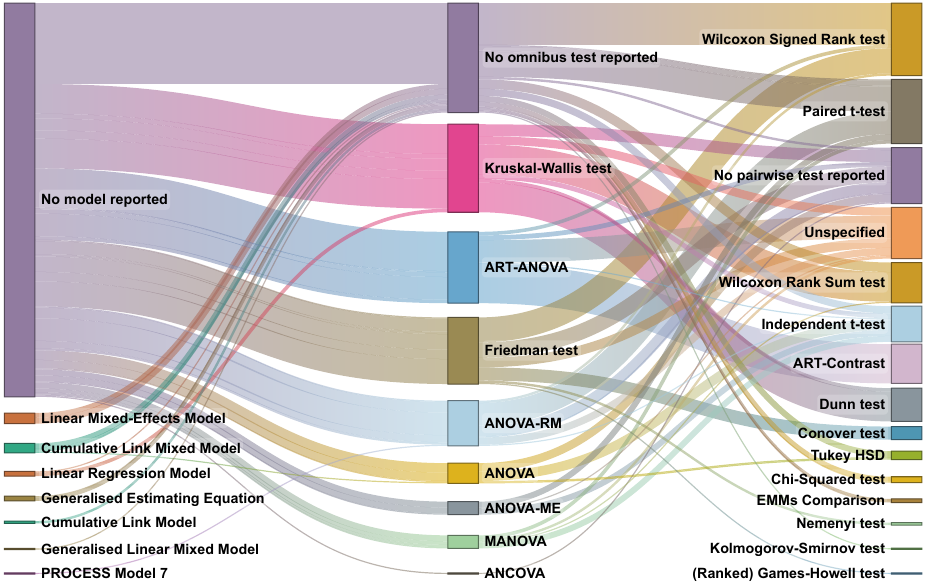}
    \caption{Sankey diagram showing the progression of statistical procedures applied to constructs measured with ordinal data. Methods are displayed in sequence from predictive models (left), to omnibus tests (centre), to pairwise tests (right). Bars labelled ‘No’ indicate constructs for which no method was reported in that category. Each node represents a statistical method and is coloured uniquely. This plot is the same as figure~\ref{fig:sankey_statistical_procedures} but with unique colours for each statistical method observed in our sample for easy identification.}
    \Description{Sankey Diagram depicting the sequence of statistical approaches used to analyse ordinal data in our sample; starting from predictive models, to omnibus tests and finally to pairwise comparisons. The Figure depicts a lack of use of predictive models to analyse ordinal data in our sample. Additionally, the high number of branching links between the categories (model/omnibus/pairwise) suggests a lack of consensus in the field. This plot depicts each statistical method observed in our sample with a unique colour for easy identification.}
    \label{fig:sankey_statistical_procedures_individual}
\end{figure*}

\clearpage

\end{document}